\newcommand*{\ATLASLATEXPATH}{./}
\documentclass[cernpreprint,texlive=2011,txfonts,UKenglish]{\ATLASLATEXPATH atlasdoc} 
\pdfoutput=1
\usepackage{\ATLASLATEXPATH atlaspackage}
\usepackage{\ATLASLATEXPATH atlasphysics} 
\usepackage{\ATLASLATEXPATH atlascover}
\usepackage{\ATLASLATEXPATH atlasbiblatex}
\usepackage{graphicx}
\usepackage{rotating}
\usepackage{multirow}
\usepackage{dcolumn}
\usepackage{xspace}
\usepackage{collref}

\usepackage{url}
\usepackage{ifpdf}
\ifpdf\pdfpageattr{/Group <</S /Transparency /I true /CS /DeviceRGB>>}\fi 

\usepackage{subfig}
\usepackage{mathrsfs}
\usepackage{accents}
\usepackage{IEEEtrantools}

\newcommand{\Z}{\ensuremath{Z}}

\newcommand{\Zp}{\ensuremath{Z_d}}
\newcommand{\HZpZp}{\ensuremath{H \rightarrow \Zp\Zp}}
\newcommand{\HZpZpllll}{\ensuremath{\HZpZp \rightarrow \llll}}
\newcommand{\HZZ}{\ensuremath{H \rightarrow \Z\Z^{*}}}
\newcommand{\HZZllll}{\ensuremath{\HZZ \rightarrow \llll}}
\newcommand{\HZZp}{\ensuremath{H \rightarrow \Z\Zp}}
\newcommand{\HZZpllll}{\ensuremath{\HZZp \rightarrow \llll}}

\newcommand{\llll}{\ensuremath{4\ell}}
\newcommand{\ZZllll}{\ensuremath{ZZ^{*} \rightarrow 4\ell}}

\newcommand{\Zdark}{\ensuremath{Z_d}}
\newcommand{\ZStar}{\ensuremath{Z^{*}}}
\newcommand{\Zstar}{\ZStar}
\newcommand{\monetwo}{\ensuremath{m_{12}}}
\newcommand{\mthreefour}{\ensuremath{m_{34}}}

\newcommand{\HZdfl}{\ensuremath{H\rightarrow Z\Zdark \rightarrow 4\ell}}
\newcommand{\HZsfl}{\ensuremath{H\rightarrow Z\ZStar \rightarrow 4\ell}}

\newcommand{\Hfl}{\ensuremath{H\rightarrow 4\ell}}

\newcommand{\relbr}{\ensuremath{R_B}}

\AtlasTitle{
Search for new light gauge bosons in Higgs boson decays to four-lepton final states in $pp$ collisions at $\sqrt{s}=8~$TeV 
with the ATLAS detector at the LHC
}

\author{The ATLAS Collaboration}

\AtlasRefCode{EXOT-2013-15}

\PreprintIdNumber{CERN-PH-EP-2015-111}

\AtlasJournal{Phys. Rev. D}

\AtlasCoverTwikiURL{}
\AtlasCoverSupportingNote{ATL-COM-PHYS-2013-1308} {http://cds.cern.ch/record/1599569}
\AtlasCoverSupportingNote{ATL-COM-PHYS-2014-115} {https://cds.cern.ch/record/1662365}

\AtlasCoverCommentsDeadline{Monday May 4, 2015}
\AtlasCoverAnalysisTeam{
K.~A.~Assamagan (*),
M.~Aurousseau,
O.~K.~Baker,
E.~Castaneda Miranda,
S.~H.~Connell,
D.~Paredes Hernandez,
T.~Lagouri (*),
C.~Petridou.
}

\AtlasCoverEdBoardMember{
G. Herten (*),
S.-C.~Hsu,
K.~Nikolopoulos,
D.~Whiteson.
}

\AtlasCoverEgroupEditors{atlas-exot-2013-15-editors@cern.ch}

\AtlasCoverEgroupEdBoard{atlas-exot-2013-15-editorial-board@cern.ch}

\AtlasAbstract{
This paper presents a search for Higgs bosons decaying to four leptons, either electrons or muons, via one or two light exotic gauge bosons $Z_d$,
$H\to Z Z_d \to 4\ell$ or \HZpZpllll{}. The search was performed using $pp$ collision data corresponding to
an integrated luminosity of about 20~\ifb\
at the center-of-mass energy of $\sqrt{s}=8~$TeV recorded with the ATLAS detector at the Large Hadron Collider.
The observed data are well described by the Standard Model prediction.
Upper bounds on the branching ratio of $\HZdfl$ and on the kinetic mixing parameter between the $Z_d$ and the Standard Model hypercharge gauge boson are set in the range
$(1$--$9)\times10^{-5}$ and $(4$--$17)\times10^{-2}$ respectively, at 95\% confidence level assuming the Standard Model branching ratio of $\HZsfl$, for $Z_d$ masses between 15 and 55~GeV. Upper bounds on the effective mass mixing parameter between the $Z$ and the $Z_d$ are also set using the branching ratio limits in the $H \to ZZ_d\to 4\ell$ search, and are in the range $(1.5$--$8.7)\times 10^{-4}$ for $15<m_{Z_d} < 35$~GeV.
Upper bounds on the branching ratio of $H\rightarrow Z_dZ_d \to 4\ell$ and on Higgs portal coupling
parameter, controlling the strength of the coupling of the Higgs
boson to dark vector bosons, are set in the range $(2$--$3)\times10^{-5}$ and $(1$--$10)\times 10^{-4}$ respectively, at 95\% confidence level  assuming the Standard Model Higgs boson production cross sections, for $Z_d$ masses between 15 and 60~GeV.

}

\clearpage

\journal{Phys. Rev. D}

\begin{document}

\nolinenumbers

\maketitle

\section{Introduction}
\label{sec:intro}

Hidden sector or dark sector states appear in many extensions to the Standard Model (SM)~\cite{Fayet:2004bw,Finkbeiner:2007kk,ArkaniHamed:2008qn,Dudas:2012t1,Curtin:2014cca,Curtin:2013fra,Davoudiasl:2013aya,BNL2012dark,wells2008find,gopalakrishna2008higgs}, to provide a candidate for the dark matter in the universe~\cite{Clowe} or to explain astrophysical observations of positron excesses~\cite{Adriani:2008zr,ATIC:2008t1,Aguilar:110.141102}. A hidden or dark sector can be introduced with an additional $U(1)_d$ dark gauge symmetry~\cite{Curtin:2014cca,Curtin:2013fra,Davoudiasl:2013aya,BNL2012dark,wells2008find,gopalakrishna2008higgs}.

In this paper, we present model-independent searches for dark sector states. We then interpret the results in benchmark  models where the dark
gauge symmetry is mediated by a dark vector boson $Z_d$. The dark sector could couple to the SM through kinetic mixing with the hypercharge gauge
boson~\cite{Galison:1983pa,Holdom:1985ag,Dienes:1996zr}. In this hypercharge portal scenario, the kinetic mixing parameter $\epsilon$ controls the coupling strength of
the dark vector boson and SM particles. If, in addition, the $U(1)_d$ symmetry is broken by the introduction of a dark Higgs boson, then there could also be a mixing between the SM Higgs boson and the dark sector Higgs boson~\cite{Curtin:2014cca,Curtin:2013fra,Davoudiasl:2013aya,BNL2012dark,wells2008find,gopalakrishna2008higgs}. In this scenario, the Higgs portal coupling $\kappa$ controls the strength of the Higgs coupling to dark vector bosons.  The observed Higgs boson would then be the lighter partner of the new Higgs doublet, and could also decay via the dark sector. There is an additional Higgs portal scenario where there could be a mass-mixing between the SM $Z$ boson and $Z_d$~\cite{Davoudiasl:2013aya,BNL2012dark}. In this scenario, the dark vector boson $Z_d$ may couple to the SM $Z$ boson with a coupling proportional to the mass mixing parameter $\delta$.  

The presence of the dark sector could be inferred either from deviations from the SM-predicted rates of Drell-Yan (DY) events or from Higgs boson decays through exotic intermediate states. Model-independent upper bounds, from electroweak constraints, on the kinetic mixing parameter of $\epsilon \leq 0.03$ are reported in
Refs.~\cite{Curtin:2014cca,Hook:2010tw,Pospelov:2008zw} for dark vector boson masses between 1~GeV and 200~GeV. Upper bounds on the kinetic mixing parameter based on searches for dilepton resonances, $pp \to\Z_d \to \ell\ell$, below the $Z$-boson mass are found to be in range of $0.005$--$0.020$ for dark vector boson masses between 20 and 80~GeV~\cite{Hoenig:2014dsa}. 
The discovery of the Higgs 
boson~\cite{englert1964,higgs1964,guralnik1964} during Run 1 of the Large Hadron Collider (LHC)~\cite{atlas2012observation,cms2012observation} opens a new and rich experimental 
program that includes the search for exotic decays \HZZpllll{} and \HZpZpllll{}. This scenario is not entirely excluded by electroweak constraints~\cite{Curtin:2014cca,Curtin:2013fra,Davoudiasl:2013aya,BNL2012dark,wells2008find,gopalakrishna2008higgs,Hook:2010tw,Hoenig:2014dsa}. The $H \to ZZ_d$ process probes the parameter space of $\epsilon$ and $m_{Z_d}$, or $\delta$  and $m_{Z_d}$, where $m_{Z_d}$ is
the mass of the dark vector boson, and the $H \to Z_dZ_d$ process covers the parameter space of $\kappa$ and $m_{Z_d}$~\cite{Curtin:2014cca,Curtin:2013fra}. 
DY production, $pp \to\Z_d \to \ell\ell$, offers the most promising discovery potential for dark vector bosons in the event of no mixing between the dark Higgs boson and the SM Higgs boson. The \HZZpllll{} process offers a discovery potential complementary to the DY
process for $m_{Z_d}<m_Z$~\cite{Curtin:2014cca,Hoenig:2014dsa}. Both of these would be needed to understand the properties of the dark vector boson~\cite{Curtin:2014cca}. If the dark Higgs boson mixes with
the SM Higgs boson, the \HZpZpllll{} process would be important, probing the dark sector through the Higgs portal coupling~\cite{Curtin:2014cca,Curtin:2013fra}.  

This paper presents a search for Higgs bosons decaying to four leptons via one or two $Z_d$ bosons using $pp$ collision data at $\sqrt{s} = 8$~TeV collected at the CERN LHC with the ATLAS experiment.  The search uses a dataset corresponding to an integrated luminosity of 20.7~fb$^{-1}$ with an uncertainty of 3.6\% for \HZZpllll{} based on the luminosity calibration used in Refs.~\cite{moriond:2013t1CONF,atlas-stat1:2013t2},  and 20.3~fb$^{-1}$ with an uncertainty of 2.8\% for \HZpZpllll{} based on a more recent calibration~\cite{Aad:2013ucp}.  Same-flavor decays of the $Z$ and $Z_d$ bosons to electron and muon pairs are considered, giving the  
$4e$, $2e2\mu$, and $4\mu$ final states. Final states including $\tau$ leptons are not considered in the \HZZpllll{} and \HZpZpllll{} decays. In the absence of a significant  signal, upper bounds 
are set on the relative branching ratios  $\mathrm{BR} (H\to ZZ_d \to 4\ell) / \mathrm{BR}(H \to 4\ell)$ and $\mathrm{BR} (H\to Z_dZ_d \to 4\ell) / \mathrm{BR}(H \to ZZ^* \to 4\ell)$  as functions of the mass of the dark vector boson $m_{Z_d}$. The branching ratio limits are used to
set upper bounds on the kinetic mixing, mass mixing, and Higgs boson mixing parameters~\cite{Curtin:2014cca,Curtin:2013fra}. The search is restricted to 
the mass range where the \Zp{} from the decay of the Higgs boson is on-shell, i.e. $15\, \mathrm{GeV} <m_{Z_d} < m_H/2$, where $m_H=125$~GeV. Dark vector boson masses below 
15~GeV are not considered in the present search. Although the low-mass region is theoretically well motivated~\cite{Davoudiasl:2013aya,BNL2012dark}, the high 
$p_{\mathrm{T}}$ of the $Z_d$ boson relative to its mass leads to signatures that are better studied in dedicated searches~\cite{Aad:2014yea}.

The paper is organized as follows. The ATLAS detector is briefly described in Sec.~\ref{sec:atlas}.  The signal and background 
modeling is summarized in Sec.~\ref{sec:mc}. The dataset, triggers, and event reconstruction are presented in 
Sec.~\ref{sec:recons}. Detailed descriptions of the searches are given in Sec.~\ref{sec:ZZdark} and Sec.~\ref{sec:ZdarkZdark} for 
$H\to ZZ_d \to 4\ell$ and \HZpZpllll{} processes respectively. Finally, the concluding remarks are presented in Sec.~\ref{sec:conc}.

\section{Experimental Setup}
\label{sec:atlas}
The ATLAS detector~\cite{Aad:2009wy} covers almost the whole solid angle around the collision point with layers of tracking detectors, 
calorimeters and muon chambers. The ATLAS inner detector (ID) has full coverage\footnote{ATLAS uses a right-handed coordinate system with 
its origin at the nominal interaction point (IP) in the center of the detector and the $z$-axis along the beam pipe. The $x$-axis points 
from the IP to the center of the LHC ring, and the $y$-axis points upward.  The azimuthal angle $\phi$ is measured around the beam axis, 
and the polar angle $\theta$ is measured with respect to the $z$-axis. ATLAS defines transverse energy $E_{\rm T} = E \ {\rm sin}  \theta$,  
transverse momentum $\pt = p \ {\rm sin} \theta$, and pseudorapidity $\eta = - {\rm ln}({\rm tan}[\theta/2)]$.}  in the azimuthal angle 
$\phi$ and covers the 
pseudorapidity range $|\eta|<2.5$. It consists of a silicon pixel detector, a silicon microstrip detector, and a straw-tube tracker that 
also measures transition radiation for particle identification, all immersed in a 2~T axial  magnetic field produced by a superconducting solenoid.

High-granularity liquid-argon (LAr) electromagnetic sampling calorimeters, with excellent energy and position resolution, cover the 
pseudorapidity range $|\eta|<$~3.2. The hadronic calorimetry in the range $|\eta|<$~1.7 is provided by a scintillator-tile calorimeter, 
consisting of a large barrel and two smaller extended barrel cylinders, one on either side of the central barrel. The LAr endcap 
($1.5<\left|\eta\right|<3.2$) and forward sampling calorimeters ($3.1<\left|\eta\right|<4.9$) provide electromagnetic and hadronic 
energy measurements.

The muon spectrometer (MS) measures the deflection of muon trajectories with $|\eta| <2.7 $ in a toroidal magnetic field. Over most of the $\eta$-range, 
precision measurement of the track coordinates in the principal bending direction of the magnetic field is provided by monitored drift tubes. Cathode strip 
chambers are used in the innermost layer for $2.0 < |\eta| < 2.7$. The muon spectrometer is 
also instrumented with dedicated trigger chambers, resistive-plate chambers in the barrel and thin-gap chambers in the end-cap, covering $|\eta| <2.4 $.

The data are collected using an online  three-level trigger system~\cite{Aad:2012xs} that selects events  of interest and reduces 
the event rate from several MHz to about 400 Hz for recording and offline processing.

\section{Monte Carlo Simulation}
\label{sec:mc}

Samples of Higgs boson production in the gluon fusion (ggF) mode, with  $H \to ZZ_d \to 4\ell$ and \HZpZpllll{}, are generated for $m_H=125$~GeV and $15<m_{Z_d}<60$~GeV (in 5~GeV steps) in 
\textsc{MadGraph}5~\cite{Alwall2011uj}  with CTEQ6L1~\cite{ct10} parton distribution functions (PDF) using  the Hidden Abelian Higgs Model (HAHM) as a benchmark signal
model~\cite{Curtin:2014cca,wells2008find,gopalakrishna2008higgs}. \textsc{Pythia8}~\cite{pythia2006,pythia2008} and
\textsc{Photos}~\cite{photos2006,photos2007,photos2010} are used to take into account parton showering, hadronization, and initial- and final-state radiation.

The background processes considered in the $H \to ZZ_d \to 4\ell$ and \HZpZpllll{} searches follow those used in the \HZZllll{} 
measurements~\cite{moriond:2013t1}, and consist of:
\begin{itemize}
\item Higgs boson production via the SM ggF, VBF (vector boson fusion), $WH$, $ZH$, and $t\bar{t}H$ processes with \HZZllll{} final states. In the \HZpZpllll{} search, these background processes are normalized with the theoretical cross sections,  
where the Higgs boson production cross sections and decay branching ratios, as well as their uncertainties, are taken from 
Refs.~\cite{LHCHiggsCrossSectionWorkingGroup:2011ti,LHCHiggsCrossSectionWorkingGroup:2012vm}. In the $H \to ZZ_d \to 4\ell$ search, the normalization of $H \to 4\ell$
 is determined from data. The cross section for the ggF process 
has been calculated to next-to-leading-order (NLO)~\cite{Djouadi:1991tka,Dawson:1990zj,Spira:1995rr} and next-to-next-to-leading-order (NNLO)~\cite{Harlander:2002wh,Anastasiou:2002yz,Ravindran:2003um} in 
QCD.  In addition, QCD soft-gluon resummations calculated in the next-to-next-to-leading-logarithmic (NNLL) approximation are applied for the ggF process~\cite{Catani:2003zt}. NLO 
electroweak (EW) radiative corrections are also applied~\cite{Aglietti:2004nj,Actis:2008ug}.  These results are compiled in 
Refs.~\cite{deFlorian:2012yg,Anastasiou:2012hx,Baglio:2010ae} assuming factorization between QCD and EW corrections. For the VBF process, 
full QCD and EW corrections up to NLO~\cite{Ciccolini:2007jr,Ciccolini:2007ec,Arnold:2008rz} and approximate NNLO QCD~\cite{Bolzoni:2010xr} corrections 
are used to calculate the cross section. The cross sections for the associated $WH$ and $ZH$ production processes are calculated at NLO~\cite{Han:1991ia} and 
at NNLO~\cite{Brein:2003wg} in QCD, and NLO EW radiative corrections are applied~\cite{Ciccolini:2003jy}.  The cross section for associated Higgs boson production with a $t\bar{t}$ pair is calculated at NLO in QCD~\cite{Beenakker:2001rj,Beenakker:2002nc,Dawson:2002tg,Dawson:2003zu}.  
The SM ggF and VBF processes producing \HZZllll{} backgrounds are modeled with \textsc{Powheg}, \textsc{Pythia8} and CT10 PDFs~\cite{ct10}. The SM $WH$, $ZH$, and $t\bar{t}H$ processes   
producing \HZZllll{} backgrounds are modeled with \textsc{Pythia8} with CTEQ6L1 PDFs.
\item SM $ZZ^{*}$ production. The rate of this  background is estimated using simulation normalized to the SM cross section at NLO. The $ZZ^{*} \rightarrow 4\ell$ 
background is modeled using simulated samples generated with \textsc{Powheg}~\cite{zzherwig2011} and \textsc{Pythia8}~\cite{pythia2008} for 
$q\bar{q} \rightarrow ZZ^{*}$, and \texttt{gg2ZZ}~\cite{gg2zz2012} and \textsc{Jimmy}~\cite{JIMMY} for $gg \rightarrow ZZ^{*}$, and CT10 PDFs for both.
\item $Z$+jets and $\ttbar$. The rates of these background processes are estimated using data-driven methods. However Monte Carlo (MC) simulation is used to understand the systematic uncertainty on the data-driven techniques. The $Z+$jets production is modeled with up to five     
partons using \textsc{Alpgen}~\cite{Mangano:2002ea} and is divided into two sources: $Z+$light-jets, which includes 
$Zc\bar{c}$ in the massless $c$-quark approximation and $Zb\bar{b}$ with $b\bar{b}$ from parton showers; and $Zb\bar{b}$ using matrix-element calculations
that take into account the $b$-quark mass.  The matching scheme of matrix elements and parton shower evolution (see Ref.~\cite{Mangano:2006rw} and
the references therein) is used to remove any double counting of identical jets produced 
via the matrix-element calculation and the parton shower, but this scheme is not implemented for $b$-jets.  Therefore, $b\bar{b}$ pairs with
separation $\Delta R\equiv\sqrt{\left(\Delta\phi\right)^2+\left(\Delta\eta\right)^2} > 0.4$ between the $b$-quarks are taken from the matrix-element 
calculation, whereas for $\Delta R<0.4$ the parton-shower $b\bar{b}$~pairs are used. For comparison between data and simulation, the NNLO QCD 
\textsc{FEWZ}~\cite{Melnikov:2006kv,Anastasiou:2003ds} and NLO QCD \textsc{MCFM}~\cite{Campbell:2010ff,Campbell:2000bg} cross-section calculations are 
used to normalize the simulations for inclusive $Z$ boson and $Zb\bar{b}$ production, respectively. The $t\bar{t}$ background is simulated with 
MC@NLO-4.06~\cite{mcatnlo2003} with parton showers and underlying-event modeling as implemented in \textsc{Herwig 6.5.20}~\cite{herwig} and 
\textsc{Jimmy}. The AUET2C~\cite{ATLAStunes} tune for the underlying events is used for $t\bar{t}$ with CT10 PDFs.
\item SM $WZ$ and $WW$ production. The rates of these backgrounds are normalized to theoretical calculations at NLO in 
perturbative QCD~\cite{Campbell:2011bn}. The simulated event samples are produced with \textsc{Sherpa}~\cite{Sherpa} and CT10 PDFs.
\item Backgrounds containing $J/\psi$ and $\Upsilon$, namely $ZJ/\psi$ and $Z\Upsilon$. These backgrounds are normalized using the 
ATLAS measurements described in Ref.~\cite{Aad:2014kba}. These processes are modeled with \textsc{Pythia8}~\cite{pythia2008} and CTEQ6L1 PDFs.
\end{itemize}

Differing pileup conditions (multiple proton-proton interactions in the same or neighboring bunch-crossings) 
as a function of the instantaneous luminosity are taken into account by overlaying simulated 
minimum-bias events generated with \textsc{Pythia8} onto the hard-scattering process and reweighting 
them according to the distribution of the mean number of interactions observed in data.
The MC generated samples are processed either with a full ATLAS detector simulation~\cite{:2010wqa} based on
the GEANT4 program~\cite{Agostinelli:2002hh} or a fast simulation based on the
parameterization of the response to the electromagnetic and hadronic
showers in the ATLAS calorimeters~\cite{FastCaloSim} and a detailed simulation of other parts of the detector and the trigger system.
The results based on the fast simulation are validated against fully simulated samples and the difference is 
found to be negligible. 
The simulated events are
reconstructed and analyzed with the same procedure as the data, using the same trigger and event 
selection criteria.

\section{Event Reconstruction}
\label{sec:recons}
A combination of single-lepton and dilepton triggers is used to select the data samples. The single-electron trigger has a transverse energy ($E_{\mathrm{T}}$) threshold of 25~\GeV{} 
while the single-muon trigger has a transverse momentum ($p_{\mathrm{T}}$) threshold of 24~\GeV. The dielectron trigger has a threshold of $E_{\mathrm{T}} = 12~\GeV$ 
for both electrons. In the case of muons, triggers with symmetric thresholds at $p_{\mathrm{T}} = 13~\GeV$ and asymmetric thresholds at 
$18~\GeV$ and $8~\GeV$ are used. Finally, electron-muon triggers are used with electron $E_{\mathrm{T}}$ thresholds of 12~GeV or 
24~\GeV{} depending on the electron identification requirement, and a muon $p_{\mathrm{T}}$ threshold of 8~\GeV. The trigger efficiency for
events passing the final selection is above 97\%~\cite{moriond:2013t1} in each of the final states considered.

Data events recorded during periods when significant portions of the relevant detector subsystems were not fully functional are rejected. These 
requirements are applied independently  of the lepton final state. Events in a time window around a noise burst in the calorimeter 
are removed~\cite{Aad:2014una}. Further, all triggered events are required to contain a reconstructed primary vertex formed from at 
least 3 tracks, each with  $p_{\mathrm{T}} > 0.4$~GeV. 

Electron candidates consist of clusters of energy deposited in the electromagnetic calorimeter and associated with ID tracks~\cite{electronperf2010}.  The 
clusters matched to tracks are required to satisfy a
set of identification criteria such that the longitudinal and transverse shower profiles are consistent
with those expected from electromagnetic showers. The electron transverse momentum is computed from
the cluster energy and the track direction at the interaction point. Selected electrons must satisfy 
$E_{\mathrm{T}} > 7~\GeV$ and $|\eta| < 2.47$. Each electron must have a longitudinal impact parameter ($z_0$) 
of less than 10~mm with respect to the reconstructed primary vertex, defined as the vertex with at least three 
associated tracks for which the $\sum p_{\mathrm{T}}^2$ of the associated tracks is the highest.
Muon candidates are formed by matching reconstructed ID tracks with either complete or partial
tracks reconstructed in the muon spectrometer~\cite{wzll2010}. If a complete track is present, the two independent momentum
measurements are combined; otherwise the momentum is measured using the ID. The muon reconstruction and identification coverage is 
extended by using tracks reconstructed in the forward region ($2.5 < |\eta| < 2.7$) of the MS, which is outside the ID coverage. In the 
center of the barrel region ($|\eta| < 0.1$), where there is no coverage from muon chambers, ID tracks with $p_{\mathrm{T}} > 15~\GeV$ are identified as muons if
their calorimetric energy deposits are consistent with a minimum ionizing particle. Only one muon per
event is allowed to be reconstructed in the MS only or identified with the calorimeter. Selected muons must satisfy $p_{\mathrm{T}} > 6~\GeV$ and $|\eta| 
< 2.7$. The requirement on the longitudinal impact parameter is the same as for electrons except for the  muons reconstructed 
in the forward region without an ID track. To reject cosmic-ray muons, the impact parameter in the bending plane ($d_0$) is required 
to be within 1 mm of the primary vertex.

In order to avoid double-counting of leptons, an overlap removal procedure is applied. If two reconstructed electron candidates share the 
same ID track or are too close to each other in $\eta$ and $\phi$ ($\Delta R < 0.1$), the one with the highest transverse energy deposit in the calorimeter 
is kept. An electron within $\Delta R = 0.2$ of a muon candidate is removed, and a calorimeter-based reconstructed 
muon within $\Delta R = 0.2$ of an electron is removed.

Once the leptons have been
selected with the aforementioned basic identification and kinematic requirements, events with at least four selected leptons are kept. All  possible
combinations of four leptons (quadruplets) containing two same-flavor, opposite-charge sign (SFOS) leptons, are made.  
The selected leptons are ordered by decreasing transverse
momentum and the three highest-$p_{\mathrm{T}}$ leptons should have respectively $p_{\mathrm{T}} > 
20~\GeV$, $p_{\mathrm{T}} > 15~\GeV$ and $p_{\mathrm{T}} > 10~\GeV$. It is then required that one (two) leptons match the single-lepton (dilepton) trigger objects. The leptons
within each quadruplet are then ordered in SFOS pairs, and denoted 1 to 4, indices 1 and 2 being for the first pair, 3 and 4 for the second pair.
	
\section{$H \to ZZ_d \to 4\ell$}
\label{sec:ZZdark}
\newlength{\dhatheight}
\newcommand{\doublehat}[1]{%
    \settoheight{\dhatheight}{\ensuremath{\hat{#1}}}%
    \addtolength{\dhatheight}{-0.05ex}%
    \hat{\vphantom{\rule{1pt}{\dhatheight}}%
    \smash{\hat{#1}}}}

\subsection{Search strategy}
\label{sec:paper1-strategy-samples}

The $\HZdfl$ search is conducted with the same sample of selected $4\ell$ events as used in Refs.~\cite{moriond:2013t1CONF,atlas-stat1:2013t2} with the four-lepton invariant mass requirement of $115 < m_{4\ell} < 130$~GeV.   
This collection of events is referred to as the $4\ell$ sample.  
The invariant mass of the opposite-sign, same-flavor pair closest to the $Z$-boson pole mass of 91.2~GeV~\cite{pdg} is denoted $\monetwo$. 
The invariant mass of the remaining dilepton pair is defined as $\mthreefour$.  
The $\Hfl$ yield, denoted $n(H \to 4\ell)$, is determined by subtracting the relevant backgrounds from the $4\ell$ sample  
as shown in Eq.~(\ref{eqn:H4l}): 

\begin{align}
\label{eqn:H4l}
  n(\Hfl) =      & n(4\ell) - n(ZZ^*) - n(t\bar{t}) - n(Z+\mathrm{jets}). 
\end{align}

The other backgrounds from $WW$, $WZ$, $ZJ/\psi$ and $Z\Upsilon$ are negligible and not considered. 

\begin{figure*}[!htbp]
\includegraphics[width=\textwidth]{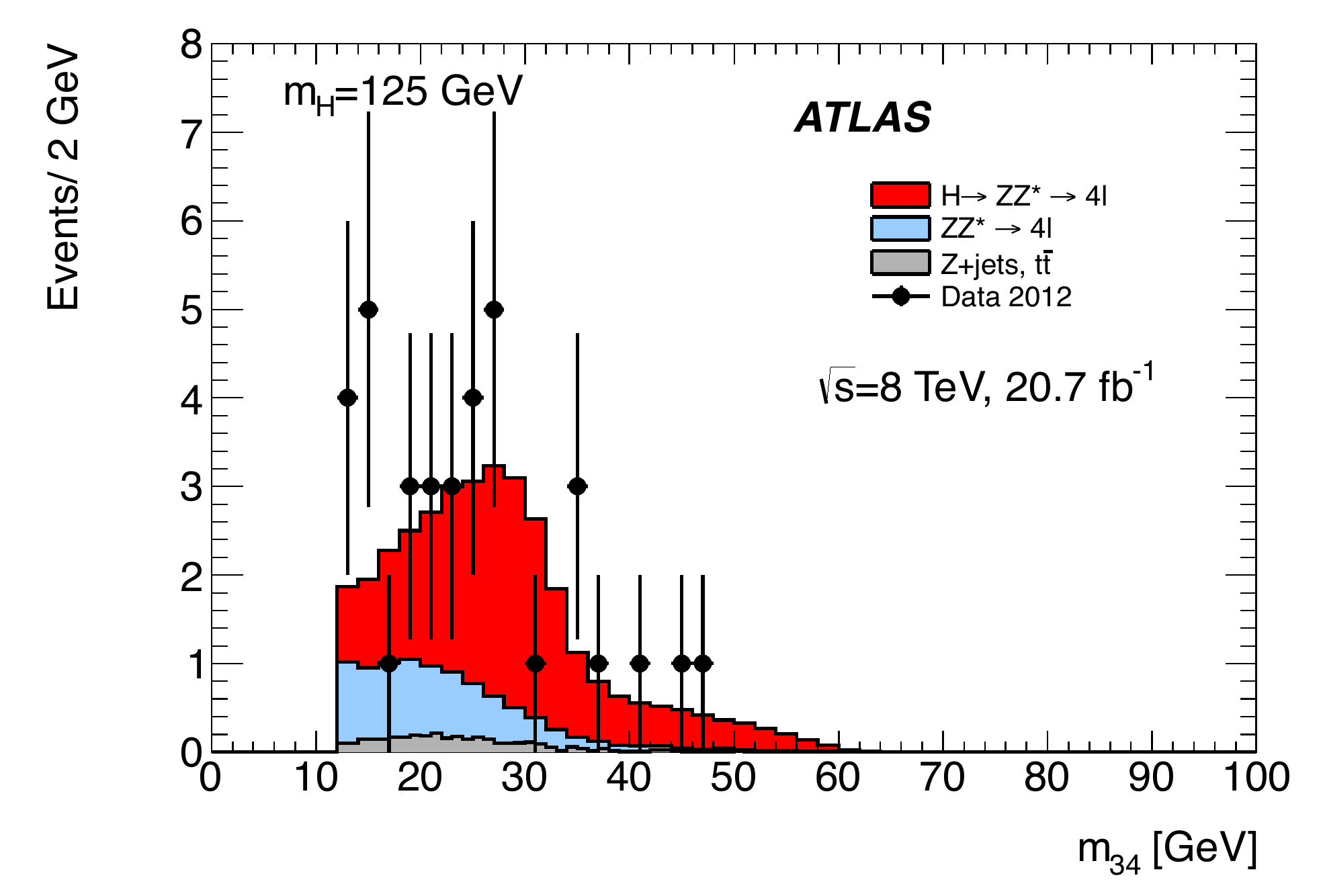}
  \caption{The distribution of the mass of the second lepton pair, $\mthreefour$, of the $\sqrt{s}$= 8 TeV data (filled circles with error bars) and the expected (pre-fit) backgrounds.
The $\HZsfl$ expected (pre-fit) normalization, for a mass hypothesis of $m_{H}$=125~GeV, is set by subtracting
the expected contributions of the $ZZ^*$, $Z$+jets and $t\bar{t}$ backgrounds from the total number of observed $4\ell$ events in the data.
}
  \label{fig:z2aftercutb-mc}
\end{figure*}

The search is performed by inspecting the $\mthreefour$ mass spectrum and testing for a local excess consistent 
with the decay of a narrow $\Zdark$ resonance.
This is accomplished through a template fit of the $\mthreefour$ distribution, using histogram-based
templates of the $\HZdfl$ signal and backgrounds. The signal template is obtained from simulation and is described in Sec.~\ref{sec:signal}. 
The $\mthreefour$ distributions and the expected normalizations of the $t\bar{t}$ and $Z$+jets backgrounds, 
along with the $\mthreefour$ distributions of the $\HZsfl$ background, as shown in Fig.~\ref{fig:z2aftercutb-mc}, 
are determined as described in Sec.~\ref{sec:backgrounds}. 
The pre-fit signal and $\HZsfl$ background event yields are set equal to the $\Hfl$ observed yield given by Eq.~(\ref{eqn:H4l}).
The expected yields for the $4\ell$ sample are shown in Table~\ref{tab:eventyields1}. 

\begin{table*}[tb]
  \begin{center}
    \begin{tabular}{lcccccc}
      \hline \hline
     Channel 
&\multicolumn{1}{c}{$ZZ^*$}
&\multicolumn{1}{c}{$t\bar{t}$ + $Z$+jets}
&\multicolumn{1}{c}{Sum}
&\multicolumn{1}{c}{Observed}
&\multicolumn{1}{c}{$\Hfl$}
\\
      \hline
      4$\mu$    & $3.1\pm 0.02\pm 0.4$  & $0.6 \pm 0.04\pm 0.2$  & $3.7 \pm 0.04\pm 0.6$  & 12  &  $8.3  \pm 0.04\pm 0.6$  \\
      4e        & $1.3\pm 0.02\pm 0.5$  & $0.8 \pm 0.07\pm 0.4$  & $2.1 \pm 0.07\pm 0.9$  &  9  &  $6.9  \pm 0.07\pm 0.9$  \\
      2$\mu$2e  & $1.4\pm 0.01\pm 0.3$  & $1.2 \pm 0.10\pm 0.4$  & $2.6 \pm 0.10\pm 0.6$  &  7  &  $4.4  \pm 0.10\pm 0.6$   \\
      2e2$\mu$  & $2.1\pm 0.02\pm 0.3$  & $0.6 \pm 0.04\pm 0.2$  & $2.7 \pm 0.10\pm 0.5$  &  8  &  $5.3  \pm 0.04\pm 0.5$   \\
      \hline
      all       & $7.8\pm 0.04\pm 1.2$  & $3.2 \pm 0.1\pm 1.0$  & $11.1 \pm 0.1\pm 1.8$ & 36  & $24.9 \pm 0.1\pm 1.8$   \\

      \hline \hline 
    \end{tabular}
  \caption{The estimated pre-fit background yields of (MC) $ZZ^*$, (data-driven) $t\bar{t}$ + $Z$+jets, their sum, the observed $4\ell$ event yield and the estimated pre-fit $\Hfl$ contribution in the $4\ell$ sample. %
The $\Hfl$ estimate in the last column is obtained as the difference between the observed event yield and the sum of the $ZZ^*$ and  $t\bar{t}$ + $Z$+jets backgrounds. The pre-fit $\HZsfl$ background and $\HZdfl$ signal events are normalized to the $\Hfl$ observed events. The uncertainties are statistical and systematic respectively. The systematic uncertainties are discussed in Section~\ref{sec:systematics}. Uncertainties on the $\Hfl$ rates do not include the statistical uncertainty from the observed number.
}
  \label{tab:eventyields1}
  \end{center}
\end{table*}

In the absence of any significant local excess, the search can be used to constrain a relative branching ratio $\relbr$, defined as:

\begin{align}
\label{eqn:ZZdRelBR}
& R_B = \frac{\mathrm{BR}(\HZdfl)}{\mathrm{BR}(\Hfl)} \nonumber \\
& = \frac{\mathrm{BR}(\HZdfl)}{\mathrm{BR}(\HZdfl) + \mathrm{BR}(\HZsfl)},
\end{align}
\noindent
where $\relbr$ is zero in the Standard Model. A likelihood function ($\mathcal{L}$) is defined as a product of Poisson probability 
densities ($\mathcal{P}$) in each bin ($i$) of the $m_{34}$ distribution, and is used to obtain a measurement of $\relbr$ :

\begin{align}
& \mathcal{L}(\rho, \mu_{H}, \nu) = \prod\limits_{i=1}^{N_{\rm bins}} \mathcal{P}(n_{i}^{\rm obs} | n_{i}^{\rm exp}) \nonumber \\
& = \prod\limits_{i=1}^{N_{\rm bins}} \mathcal{P}(n_{i}^{\rm obs} | \mu_{H} \times ( n_{i}^{Z^*} + \rho \times n_{i}^{Z_d}) + b_{i}(\nu)),  %
\label{eqn:function}
\end{align} 
\noindent
where $\mu_{H}$ is the normalization of the $\HZsfl$ background (and allowed to float in the fit), $\rho$ the parameter of interest related to the $\HZdfl$ normalization and $\rho\times\mu_H$ the normalization  of the $\HZdfl$ signal. The symbol $\nu$ represents the systematic uncertainties on the background estimates that are treated as nuisance parameters, and $N_{\rm bins}$ the total number
of bins of the $m_{34}$ distribution. 
The likelihood to observe the yield in some bin, $n_{i}^{\rm obs}$, given the expected yield $n_{i}^{\rm exp}$ is then a function of the
 expected yields $n(\Hfl)$ of $\HZdfl$ ($\mu_{H} \times \rho \times n_{i}^{Z_d}$) and $\HZsfl$ ($\mu_{H} \times n_{i}^{Z^*}$), and the 
contribution of backgrounds $b_i(\nu)$.

An upper bound on $\rho$ is obtained from the binned likelihood fit to the data, and used in Eq.~(\ref{eqn:ZZdRelBR}) to obtain a measurement of $\relbr$, taking into account the detector 
acceptance ($\mathcal{A}$) and reconstruction efficiency ($\varepsilon$):

\begin{align}
R_{B} & = \frac{\rho\times\mu_H\times n(\Hfl)} {\rho\times\mu_H\times n(\Hfl) + C\times\mu_H\times n(\Hfl)} \nonumber \\
& = \frac{\rho}{\rho+C},  
\label{eqn:relbr1}
\end{align}
\noindent
\noindent
where $C$ is the ratio of the products of the acceptances and reconstruction efficiencies in $H\rightarrow ZZ_d\rightarrow 4\ell$ 
and $H\rightarrow ZZ^* \rightarrow 4\ell$ events:
\begin{equation}
\label{ZZd:acc}
  C = \frac{\mathcal{A}_{Z\Zdark}\times\varepsilon_{Z\Zdark}}{\mathcal{A}_{Z\Zstar}\times\varepsilon_{Z\Zstar}}.
\end{equation}
\noindent
The acceptance is defined as the fraction of 
generated events that are within a fiducial region. The reconstruction efficiency is defined as the fraction of 
events within the fiducial region that are reconstructed and selected as part of the $4\ell$ signal sample.

\subsection{Signal modeling}
\label{sec:signal}

A signal would produce a narrow peak in the \mthreefour{} mass spectrum.  
 The width of the \mthreefour{} peak for the $\Zdark$ signal is dominated by detector resolution for all $\Zdark$ masses 
considered.  
For the individual decay channels and their combination, the resolutions of the $m_{34}$ distributions are determined from Gaussian fits. 
The $m_{34}$ resolutions show a linear trend between $m_{Z_d}=15~\GeV$ and $m_{Z_d}=55~\GeV$ and vary from 0.3~\GeV{} to 1.5~\GeV{} respectively, for the combination of all the
final states.  
The resolutions of the $m_{34}$ distributions are smaller than the mass spacing between the generated signal samples 
(5~\GeV), requiring an interpolation to probe intermediate values of $m_{Z_d}$.  
Histogram-based templates are used to model the $\Zdark$ signal where no simulation is available; these templates are obtained from morphed signals 
 produced with the procedure defined in Ref.~\cite{atlasTool:1999}. The morphed signal templates are generated with a mass
spacing of 1~\GeV{}. %

The acceptances and reconstruction efficiencies of the $\HZdfl$ signal and $\HZsfl$ background are used in Eqs.~(\ref{eqn:relbr1}) and~(\ref{ZZd:acc}) to obtain the measurement of the relative branching ratio $R_B$. 
The acceptances and efficiencies are derived with $\HZdfl$ and $\HZsfl$ MC samples where
the Higgs boson is produced via ggF. The product of acceptance and reconstruction efficiency for 
VBF differs from ggF by only 1.2\% and the contribution of $VH$ and $t\bar{t}H$ production modes is negligible: the products of 
acceptance and reconstruction efficiency obtained using the ggF production mode are used also for VBF, $VH$ and $t\bar{t}H$.
\subsection{Event selection}
\label{sec:selection}

 The Higgs boson candidate is formed by selecting two pairs of SFOS leptons. 
The value of \monetwo{} is required to be between 50 GeV and 106 GeV. 
The value of \mthreefour{} is required to be in the range 12 GeV $\le \mthreefour \le$ 115 \GeV.
The four-lepton invariant mass $m_{4\ell}$ is required to be in the range $115 < m_{4\ell} < 130$~GeV. 
After applying the selection to the 8~\TeV{} data sample, 36 events are left as shown in Table~\ref{tab:eventyields1}.  
The events are grouped into four channels based on the flavor of the reconstructed leptons.  
Events with four electrons are in the $4e$ channel.  Events in which the $Z$ boson is reconstructed
with electrons, and \mthreefour{} is formed from muons, are in the $2e2\mu$ channel.  Similarly, events
in which the $Z$ is reconstructed from muons and \mthreefour{} is formed from electrons are in the
$2\mu2e$ channel.  Events with four muons are in the $4\mu$ channel.
\subsection{Background estimation}
\label{sec:backgrounds}

The search is performed using the same background estimation strategy as the $\HZsfl$ 
measurements. The expected rates of the $t\bar{t}$ and $Z$+jets backgrounds are 
estimated using data-driven methods as described in detail in Refs.~\cite{moriond:2013t1CONF,atlas-stat1:2013t2}. The results of 
the expected $t\bar{t}$ and $Z$+jets background estimations from data control regions are summarized in 
Table~\ref{tab:ddbackgrounds}. In the ``$m_{12}$ fit method'', the $m_{12}$ distribution of $t\bar{t}$ is fitted with a second-order Chebychev
polynomial, and the $Z$+jets component is fitted with a Breit-Wigner lineshape convolved with a Crystal Ball resolution function~\cite{moriond:2013t1CONF}. 
In the ``$\ell\ell+e^{\pm}e^{\mp}$ relaxed requirements'' method, a background control region is formed by relaxing the electron selection criteria 
for electrons of the subleading pairs~\cite{moriond:2013t1CONF}.
\begin{table*}[tb]
  \begin{center}
    \begin{tabular}{lcc}
      \hline \hline
     Method
&\multicolumn{1}{c}{Estimated background}
\\
      \hline
       & 4$\mu$   \\
       \hline
       $m_{12}$ fit: $Z$+jets contribution & $2.4 \pm 0.5 \pm 0.6$ \\
       $m_{12}$ fit: $\ttbar$ contribution & $0.14 \pm 0.03 \pm 0.03$ \\
       \hline
       & $2e2\mu$ \\
       \hline
       $m_{12}$ fit: $Z$+jets contribution & $2.5 \pm 0.5 \pm 0.6$ \\
       $m_{12}$ fit: $\ttbar$ contribution & $0.10 \pm 0.02 \pm 0.02$ \\
       \hline
       & $2\mu 2e$ \\
       \hline
       $\ell\ell+e^{\pm}e^{\mp}$ relaxed requirements: sum of $Z$ + jets and $\ttbar$ contributions & $5.2 \pm 0.4 \pm 0.5$ \\
       \hline
       & $4e$ \\
       \hline
       $\ell\ell+e^{\pm}e^{\mp}$ relaxed requirements: sum of $Z$ + jets and $\ttbar$ contributions & $3.2 \pm 0.5 \pm 0.4$ \\
      \hline \hline
    \end{tabular}
  \caption{Summary of the estimated expected numbers of $Z$+jets and $\ttbar$ background events for the 20.7~$\ifb$ of data at $\sqrt{s}$ = 8~TeV for the full mass range after kinematic selections, for the $H\to ZZ_d \to 4\ell$ search. The first uncertainty is statistical while the second is systematic. The uncertainties are given on the event yields. 
Approximately 80\% of the $t\bar{t}$ and $Z$+jets backgrounds have $m_{4\ell} < 160$~GeV.} 
  \label{tab:ddbackgrounds}
  \end{center}
\end{table*}
Since a fit to the data using \mthreefour{} background templates is carried out in the search, both 
the distribution in $m_{34}$ and normalization of the backgrounds are relevant.  
For all relevant backgrounds ($\HZsfl$, $ZZ^*$, $t\bar{t}$ and $Z$+jets) the $m_{34}$ distribution is obtained from simulation.

\subsection{Systematic uncertainties}
\label{sec:systematics}

The sources of the  systematic uncertainties in the $\HZdfl$ search are the same as in the $\HZsfl$ measurements. 
Uncertainties on the lepton reconstruction and identification
efficiencies, as well as on the energy and momentum reconstruction
and scale are described in detail in Refs.~\cite{moriond:2013t1CONF,atlas-stat1:2013t2}, and shown in Table~\ref{tab:ZZd_syst}.  
The lepton identification is the dominant 
contribution to the systematic uncertainties on the $ZZ^*$ background.
The largest uncertainty in the $\HZdfl$ search is the normalization of the $t\bar{t}$ and $Z$+jets backgrounds. 
Systematic uncertainties related to the 
determination of selection efficiencies of isolation and
impact parameters requirements are shown to be negligible in
comparison with other systematic uncertainties. 
The uncertainty in luminosity~\cite{Aad:2013ucp} is applied to the $ZZ^*$ background normalization. 
\begin{table*}[tb]
  \begin{center}
    \begin{tabular}{lcccc}
      \hline \hline
       \multicolumn{5}{c}{Systematic Uncertainties (\%)} \\
      \hline \hline
      Source &  4$\mu$ & 4e & 2$\mu$2e & 2e2$\mu$ \\
      \hline
       Electron Identification              &  -- & 9.4 & 8.7 & 2.4 \\
       Electron Energy Scale                &  -- & 0.4 & -- & 0.2 \\
       Muon Identification                  & 0.8 & -- & 0.4 & 0.7 \\
       Muon Momentum Scale                  & 0.2 & -- & 0.1 & -- \\
       Luminosity                           & 3.6 & 3.6 & 3.6 & 3.6 \\
       $t\bar{t}$ and $Z$+jets Normalization & 25.0 & 25.0 & 25.0 & 25.0 \\
       $ZZ^*$ (QCD scale)                   & 5.0 & 5.0 & 5.0 & 5.0 \\
       $ZZ^*$ ($\qqbar$/PDF and $\alpha_{\mathrm{S}}$) & 4.0 & 4.0 & 4.0 & 4.0 \\
       $ZZ^*$ (gg/PDF and $\alpha_{\mathrm{s}}$) & 8.0 & 8.0 & 8.0 & 8.0 \\
      \hline \hline
    \end{tabular}
  \caption{The relative systematic uncertainties on the event yields in the $\HZdfl$ search.}
  \label{tab:ZZd_syst}
  \end{center}
\end{table*}
The electron energy scale uncertainty is determined from $Z\rightarrow ee$ samples and for energies below 15 GeV from $J/\psi \rightarrow ee$
decays~\cite{moriond:2013t1CONF,atlas-stat1:2013t2}.  
Final-state QED radiation modeling and background contamination affect
the mass scale uncertainty negligibly. 
The muon momentum scale systematic uncertainty 
is determined from $Z\rightarrow \mu\mu$ samples  
and from $J/\psi \rightarrow \mu\mu$ as well as $\Upsilon \rightarrow \mu\mu$
decays~\cite{moriond:2013t1CONF,atlas-stat1:2013t2}.  
Theory related systematic uncertainties on the Higgs production cross section and branching ratios are
discussed in Refs.~\cite{LHCHiggsCrossSectionWorkingGroup:2011ti,LHCHiggsCrossSectionWorkingGroup:2012vm,moriond:2013t1}, but do not apply in this search since 
the $\Hfl$ normalization is obtained from data.  Uncertainties on the \mthreefour{} shapes arising from theory uncertainties on the PDFs and renormalization and factorization scales are found to be negligible.  Theory cross-section uncertainties  are applied to the $ZZ^*$ background. 
Normalization uncertainties  are taken into account for the $Z$+jets and $\ttbar$ backgrounds  based on the data-driven  determination of
these backgrounds. 
\subsection{Results and interpretation}
\label{sec:paper1-results-interpretation}

A profile-likelihood test statistic is used with the $CL_s$ modified frequentist formalism~\cite{CLs,Cousins:1991,2011EPJC...71.1554C,eCowan} implemented in the \verb.RooStats. 
framework~\cite{Gumpert:2014kea} to test whether the data are compatible with the signal-plus-background and background-only hypotheses. 
Separate fits are performed for different $m_{Z_d}$ hypotheses from 15~\GeV{} to 55~\GeV{}, with 1~\GeV{} spacing.
After scanning the \mthreefour{} mass spectrum for an excess consistent with the
presence of a $\HZdfl$ signal, no significant deviation from SM expectations
is observed.  

\begin{figure*}[htbp]
\includegraphics[width=\textwidth]{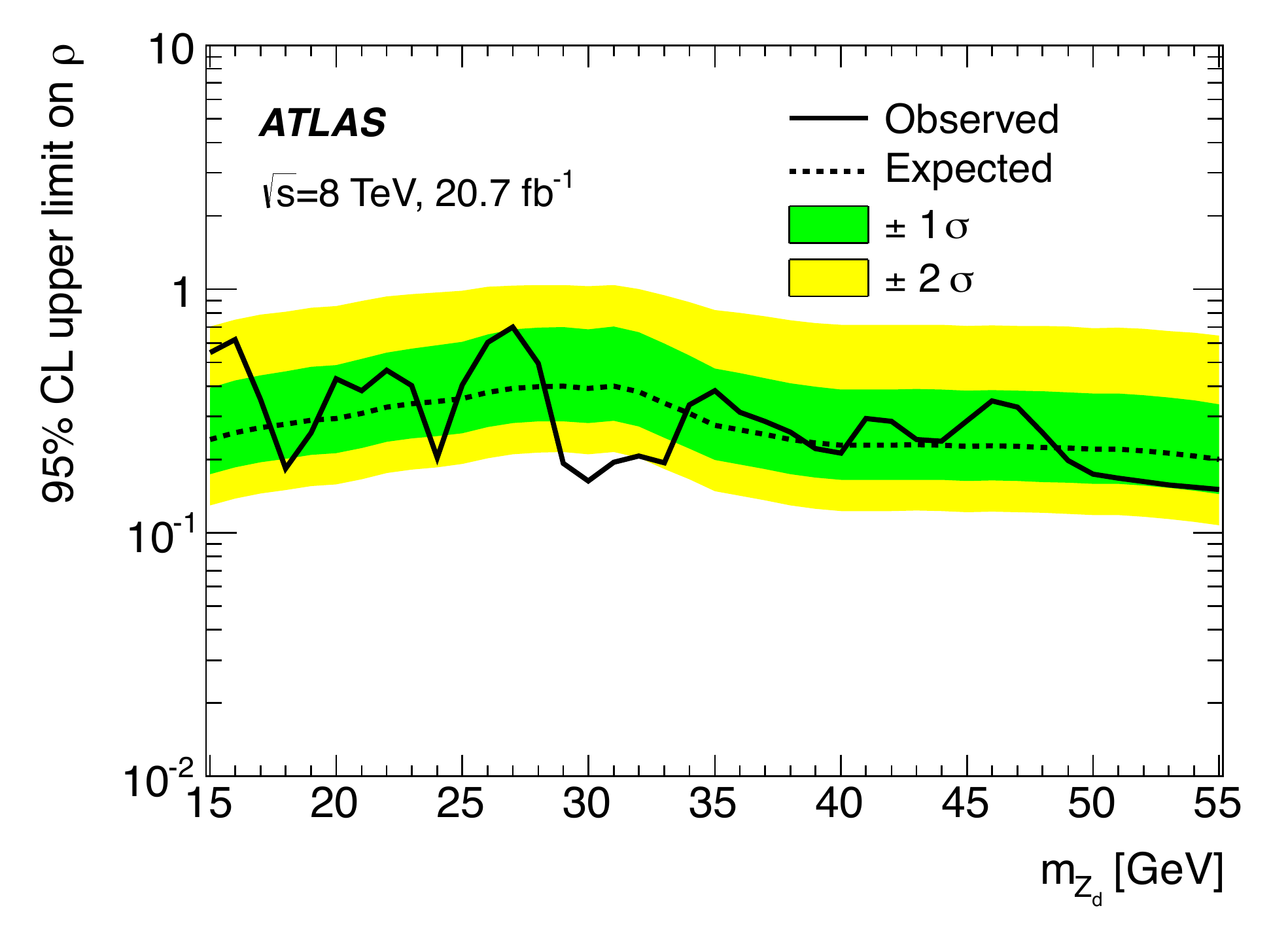}
  \caption{The observed (solid line) and median expected (dashed line) 95\% confidence level (CL) upper limits on the parameter $\rho$ related to the $H \to ZZ_d \to 4\ell$ normalization as a function of $m_{Z_d}$, %
for the combination of all four channels (4$\mu$, 4e, 2$\mu$2e, 2e2$\mu$). 
The $\pm 1\sigma$ and $\pm 2\sigma$ expected exclusion regions are indicated in green and yellow, respectively.}
  \label{fig:obs-exclim-MCZp-1}
\end{figure*}

\begin{figure*}[htbp]
\includegraphics[width=\textwidth]{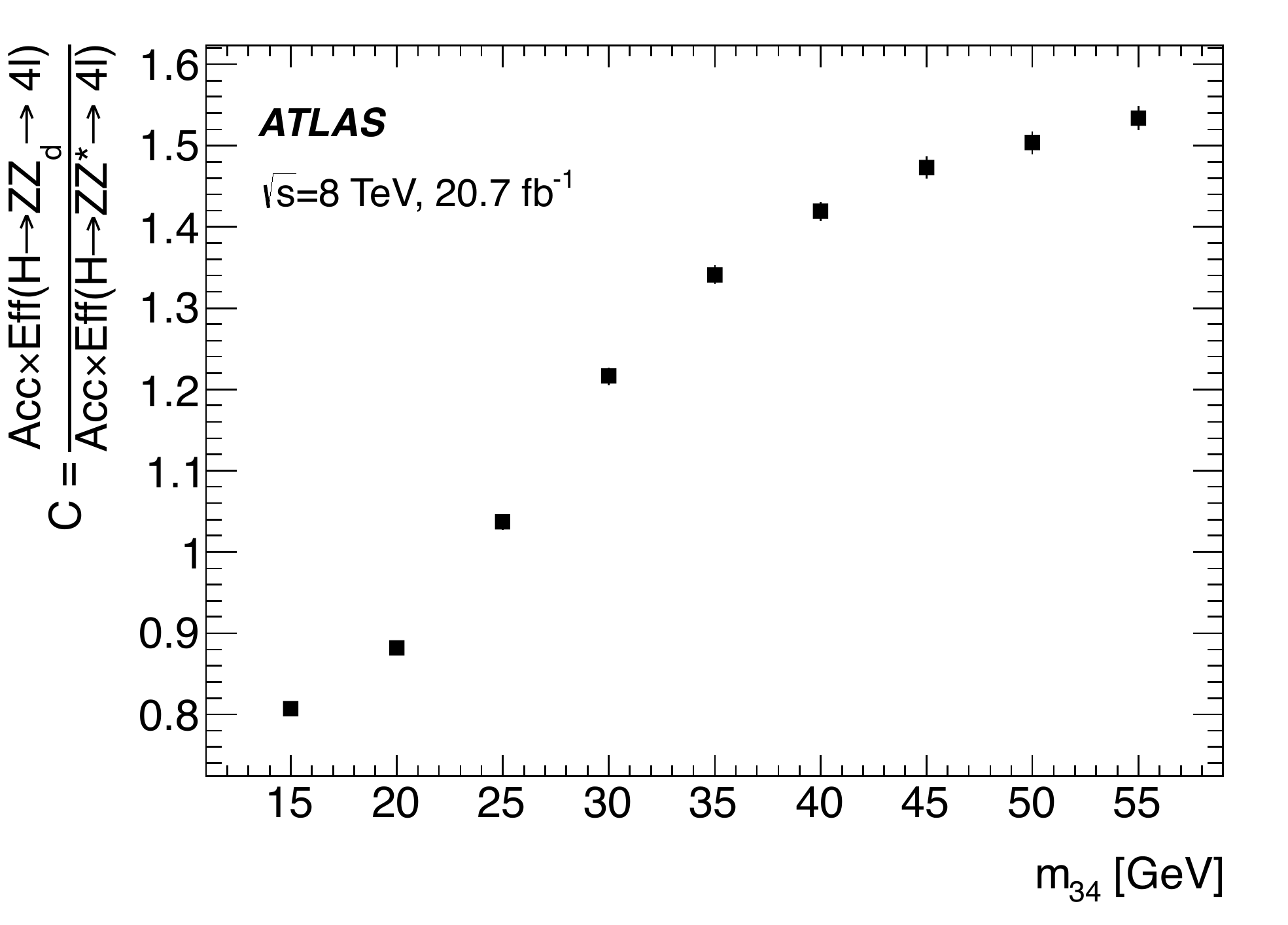}\caption{The ratio $C$ of the products of the acceptances and reconstruction efficiencies in $H\rightarrow ZZ_d\rightarrow 4\ell$ and $H\rightarrow ZZ^* \rightarrow 4\ell$ events as a function of $m_{34}$.\label{fig:effratios}}
\end{figure*}

\begin{figure*}[htbp]
\includegraphics[width=\textwidth]{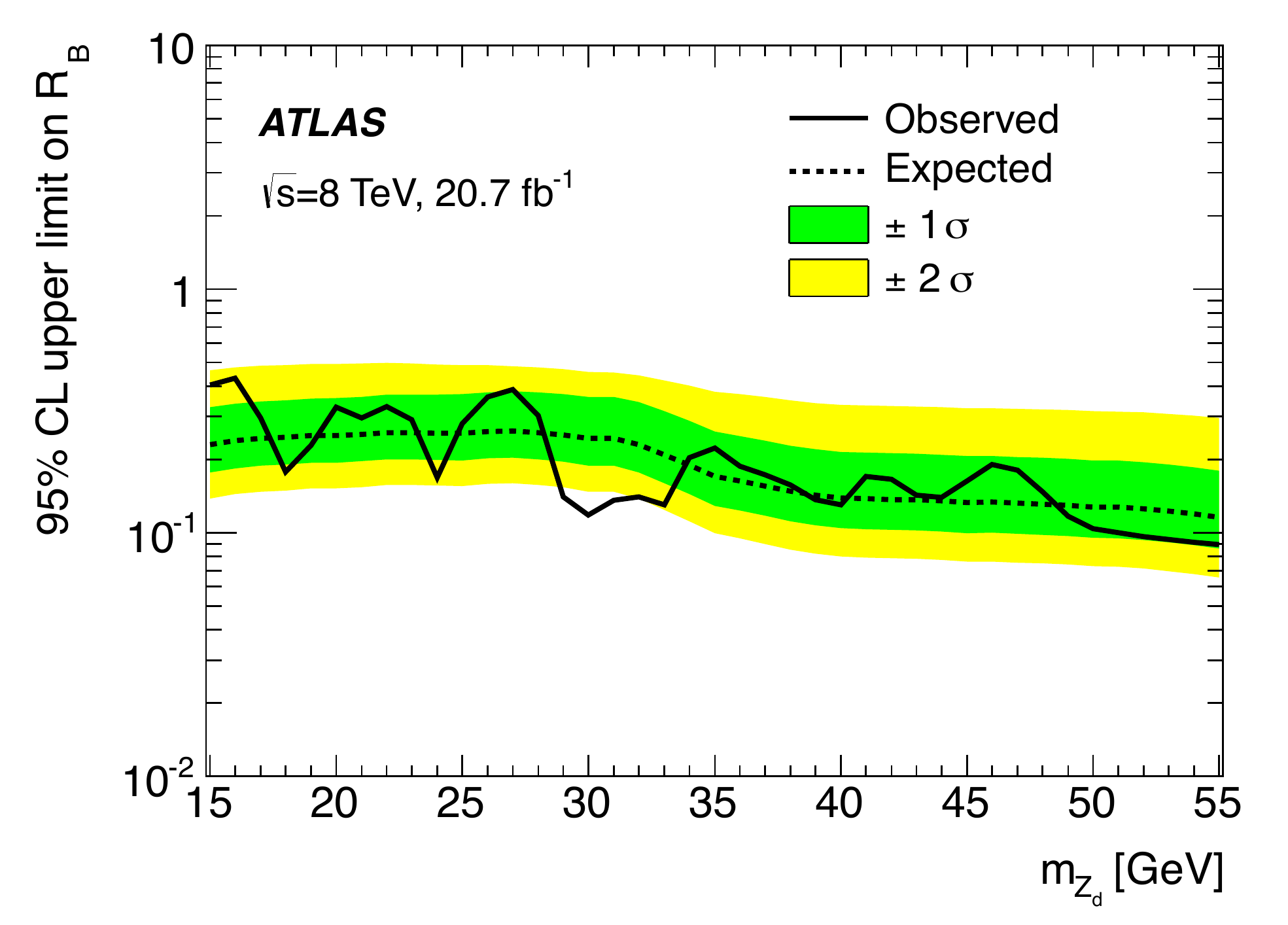}
\caption{The 95\% CL upper limits on the relative branching ratio, $\relbr=\frac{\mathrm{BR}(\HZdfl)} {\mathrm{BR}(\Hfl)}$ as a function of $m_{Z_d}$. The $\pm 1\sigma$ and $\pm 2\sigma$ expected exclusion regions are indicated in green and yellow, respectively.\label{fig:limitRelBR}}
\end{figure*}

The asymptotic approximation~\cite{2011EPJC...71.1554C} is used to estimate the expected and observed exclusion limits on $\rho$ for  
the combination of all the final states, and the result is shown in Fig.~\ref{fig:obs-exclim-MCZp-1}.
The relative branching ratio $R_B$ as a function of 
$m_{Z_d}$ is extracted using Eqs.~(\ref{eqn:ZZdRelBR}) and~(\ref{eqn:relbr1}) where the value of $C$ as a function of $\mthreefour$ is shown in Fig.~\ref{fig:effratios}, for the combination of all four final states. This is then used with $\rho$ to constrain the 
value of $\relbr$, and the result is shown in Fig.~\ref{fig:limitRelBR} for the combination of all four final states.


\begin{figure*}[htbp]
\includegraphics[width=\textwidth]{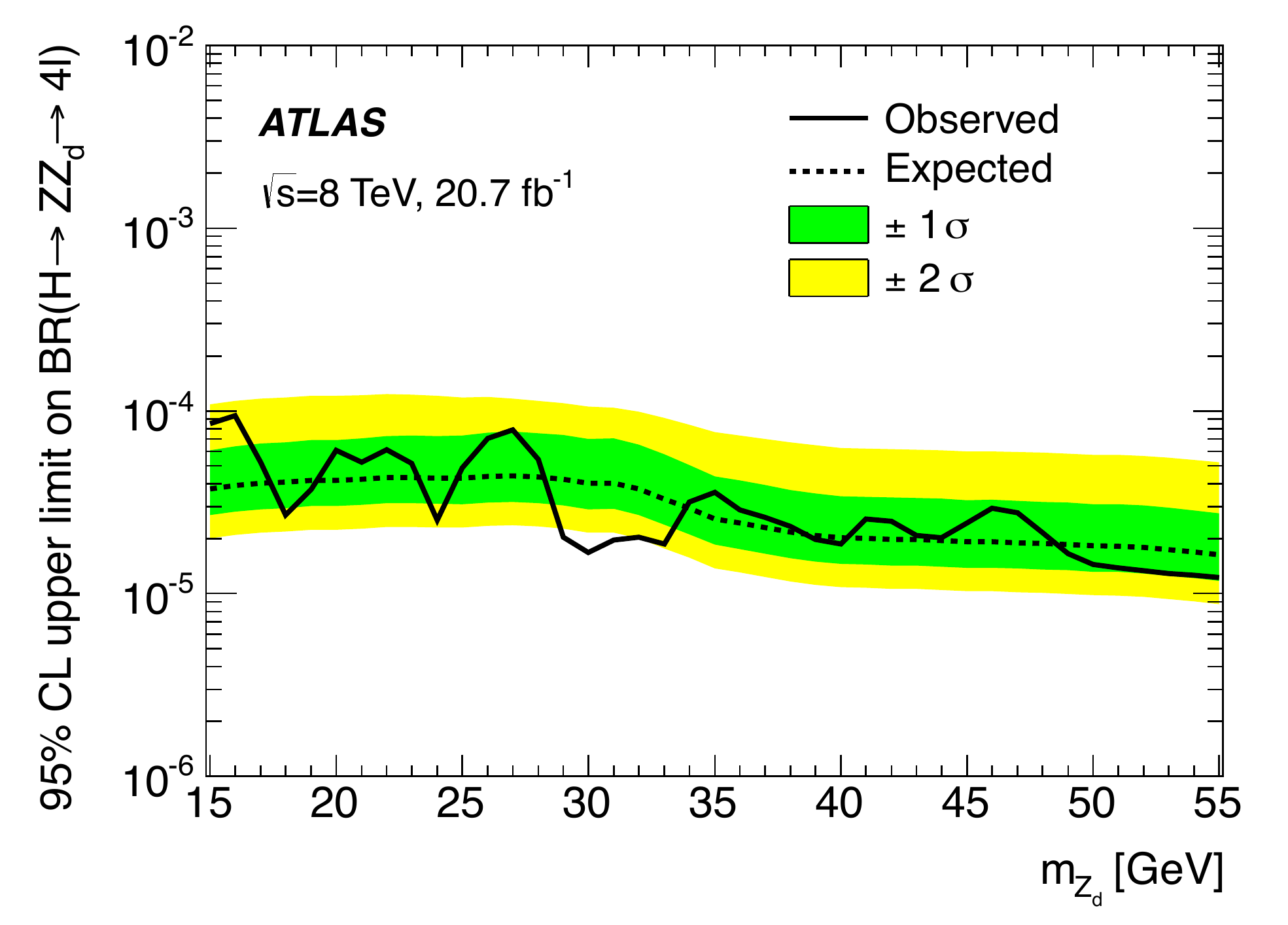}
\caption{The 95\% CL upper limits on the branching ratio of $\HZdfl$ as a function of $m_{Z_d}$ using the combined upper
limit on $\relbr$ and  the SM branching ratio of $\HZsfl$. 
\label{fig:BrZZd4l}}
\end{figure*}

\begin{figure*}[htbp]
\includegraphics[width=\textwidth]{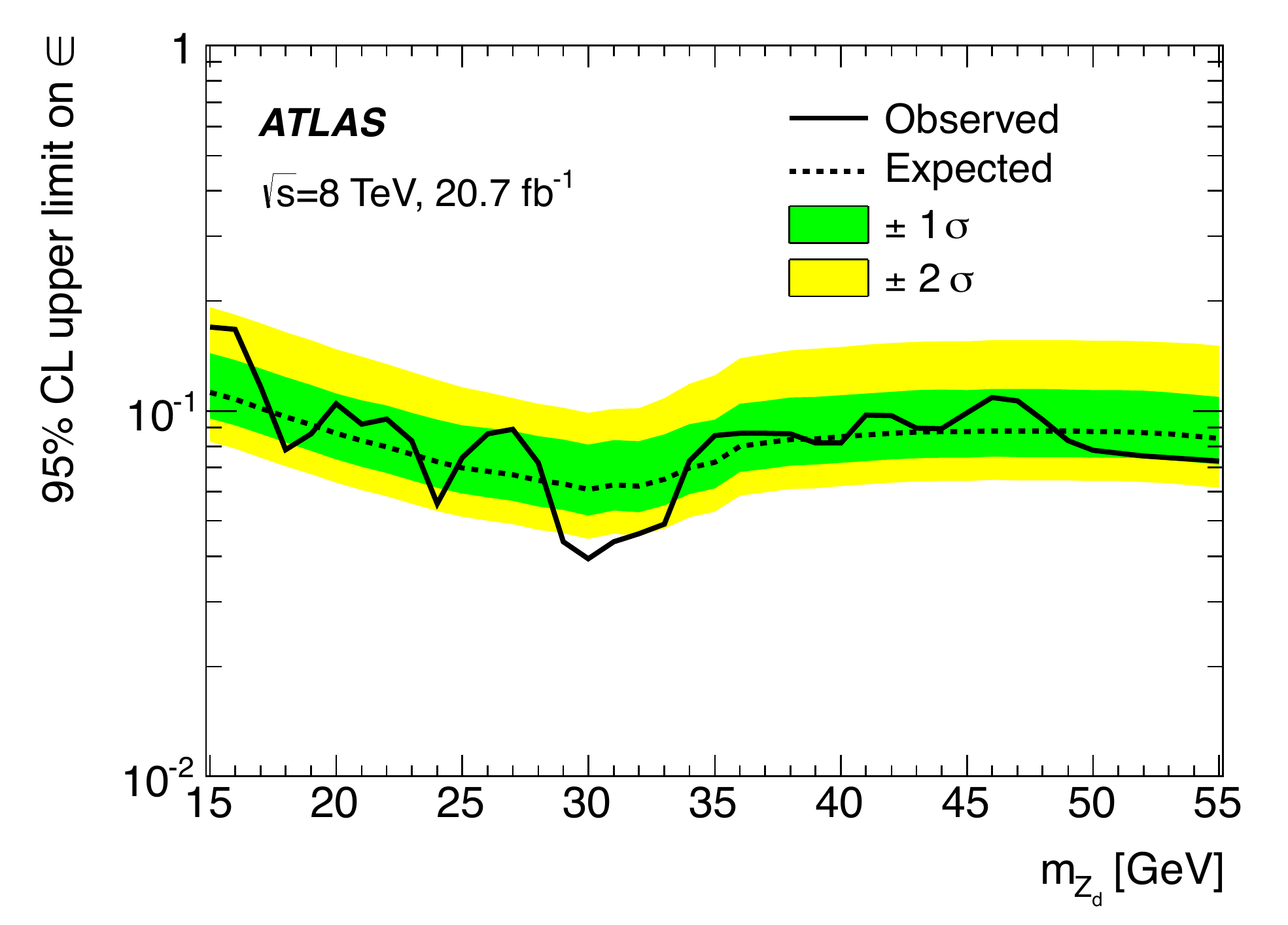}
\caption{The 95\% CL upper limits on the gauge kinetic mixing parameter $\epsilon$ as a function of $m_{Z_d}$ using the combined upper
limit on the branching ratio of $\HZdfl$ and Table~2 of Ref.~\cite{Curtin:2014cca}.
\label{fig:epsilon}}
\end{figure*}

The simplest benchmark model adds to the SM Lagrangian~\cite{Curtin:2013fra,gopalakrishna2008higgs,Davoudiasl:2013aya,BNL2012dark} a  $U(1)_d$ gauge symmetry that introduces the dark vector boson $Z_d$.
 The dark vector boson may mix kinetically with the SM hypercharge gauge boson with kinetic mixing parameter $\epsilon$~\cite{Curtin:2013fra,gopalakrishna2008higgs}.
 This enables the decay  $H\rightarrow ZZ_d$ through the Hypercharge Portal. 
 The $Z_d$ is assumed to be narrow and on-shell. Furthermore, the present search assumes prompt $Z_d$ decays consistent with current bounds on $\epsilon$
from electroweak constraints~\cite{Hook:2010tw,Pospelov:2008zw}. The coupling of the $Z_d$ to SM fermions is given in Eq.~(47) of
 Ref.~\cite{Curtin:2013fra} to be linear in $\epsilon$, so that $\mathrm{BR}(Z_d\rightarrow\ell\ell)$
 is independent of $\epsilon$ due to cancellations~\cite{Curtin:2013fra}. 
In this model, the $H \to ZZ_d \to 4\ell$ search can be used to constrain the hypercharge kinetic mixing parameter $\epsilon$ as follows:
the upper limit on $\relbr$ shown in Fig.~\ref{fig:limitRelBR} leads to an upper limit on
$\mathrm{BR}(H\rightarrow ZZ_d\rightarrow 4\ell)$ assuming the SM branching ratio of $\HZsfl$ of
 $1.25\times10^{-4}$~\cite{LHCHiggsCrossSectionWorkingGroup:2011ti,LHCHiggsCrossSectionWorkingGroup:2012vm} as shown in Fig.~\ref{fig:BrZZd4l}.
The limit on $\epsilon$ can be obtained directly from the $\mathrm{BR}(H\rightarrow ZZ_d\rightarrow 4\ell)$ upper bounds and by using Table~2 of
Ref.~\cite{Curtin:2014cca}. The 95\% CL upper bounds on $\epsilon$ are shown in Fig.~\ref{fig:epsilon} as a function of $m_{Z_d}$ in the case 
$\epsilon$~$\gg$~$\kappa$ where $\kappa$ is the Higgs portal coupling.

\begin{figure*}[htbp]
\includegraphics[width=\textwidth]{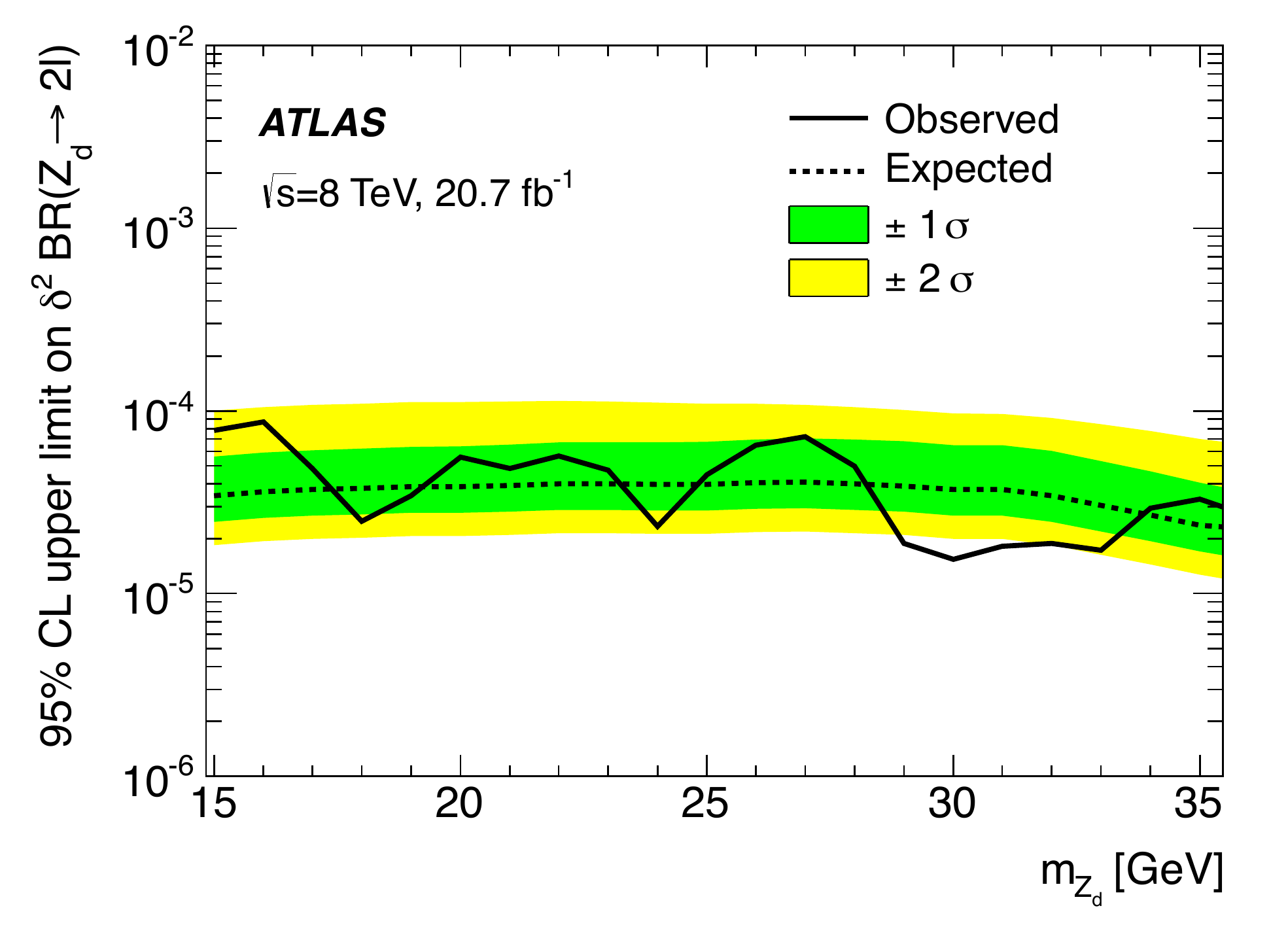}
\caption{The 95\% CL upper limits on the product of the mass-mixing parameter $\delta$ and the branching ratio of $Z_d$ decays to two leptons (electrons, or muons), $\delta^2\times \mathrm{BR}(Z_d \to 2\ell)$, as a function of $m_{Z_d}$ using the combined upper
limit on the relative branching ratio of $\HZdfl$ and the partial width of $H \to ZZ_d$ computed in Refs.~\cite{Davoudiasl:2013aya,BNL2012dark}.
\label{fig:delta}}
\end{figure*}

The measurement of the relative branching ratio $R_B$ as shown in Fig.~\ref{fig:limitRelBR} can also be used to constrain the mass-mixing parameter of the model described 
in Refs.~\cite{Davoudiasl:2013aya,BNL2012dark} where the SM is extended with a dark vector boson and another Higgs doublet, and a mass
mixing between the dark vector boson and the SM $Z$ boson is introduced. This model explores how a $U(1)_d$ gauge interaction in the hidden sector may manifest itself in the decays of 
the Higgs boson. The model also assumes that the $Z_d$, being in the hidden sector, does not couple directly to any SM particles including the Higgs boson (i.e. the SM particles do not carry dark charges). However, particles in the 
extensions to the SM, such as a second Higgs doublet, may carry dark charges allowing for indirect couplings via the $Z$-$Z_d$ mass
mixing.  The possibility of mixing between the SM Higgs boson with other scalars such
the dark sector Higgs boson is not considered for simplicity. The $Z$-$Z_d$ mass-mixing scenario also leads to potentially observable $H \to ZZ_d \to 4\ell$ decays
at the LHC even with the total integrated luminosity collected in Run~1. The partial widths of $H \to ZZ_d \to 4\ell$ and $H \to ZZ_d$ are given in terms of the $Z$-$Z_d$ mass-mixing parameter $\delta$ and $m_{Z_d}$ in Eq.~(34) of Ref.~\cite{BNL2012dark} and Eq.~(A.4) of Ref.~\cite{Davoudiasl:2013aya} respectively. As a result, using the measurement
of the relative branching ratio $R_B$ described in this paper, one may set upper bounds on the product $\delta^2\times \mathrm{BR}(Z_d \to 2\ell)$ as function of $m_{Z_d}$ as follows. From Eq.~(\ref{eqn:ZZdRelBR}) and for $m_{Z_d} < (m_H - m_Z)$

\begin{align}
& \frac{\mathrm{BR}(H \to ZZ_d\to 4\ell)}{\mathrm{BR}(H\to ZZ^*\to 4\ell)} = \frac{R_B}{(1-R_B)}, \nonumber \\
 & \simeq \frac{\Gamma(H\to ZZ_d)}{\Gamma_{\mathrm{SM}}} \times \nonumber \\ 
 & \frac{\mathrm{BR}(Z^*\to 2\ell)\times \mathrm{BR}(Z_d \to 2\ell)}{\mathrm{BR}(H\to ZZ^*\to 4\ell)}, 
\label{eqn:rbgamma}
\end{align} 
where $\Gamma_{\mathrm{SM}}$ is the total width of the SM Higgs boson and $\Gamma(H\to ZZ_d) \ll \Gamma_{\mathrm{SM}}$. From Eqs.~(4), (A.3) and (A.4) of Ref.~\cite{Davoudiasl:2013aya}, $\Gamma(H\to ZZ_d) \sim \delta^2$. It therefore follows from Eq.~(\ref{eqn:rbgamma}), with the further assumption $m^2_{Z_d}\ll (m^2_H-m^2_Z)$ that:
\begin{eqnarray}
\label{eqn:bnllink}
\frac{R_B}{(1-R_B)} & \simeq & \delta^2\times \mathrm{BR}(Z_d \to 2\ell) \times \nonumber \\
                    &        & \frac{\mathrm{BR}(Z^*\to 2\ell)}{\mathrm{BR}(H\to ZZ^*\to 4\ell)}\times \frac{f(m_{Z_d})}{\Gamma_{\mathrm{SM}}},\nonumber \\
f(m_{Z_d}) & = & \frac{1}{16\pi}\frac{(m^2_H-m^2_Z)^3}{v^2m^3_H}.
\end{eqnarray}
\noindent
where $v$ is the vacuum expectation value of the SM Higgs field. The limit is set on the product $\delta^2\times \mathrm{BR}(Z_d \to 2\ell)$ since both $\delta$ and $\mathrm{BR}(Z_d \to 2\ell)$ are model-dependent: in the case where
kinetic mixing dominates, $\mathrm{BR}(Z_d \to 2\ell) \sim 30\%$ for the model presented in Ref.~\cite{Curtin:2013fra} but it could be smaller when $Z$-$Z_d$ mass mixing dominates~\cite{BNL2012dark}. In the $m_{Z_d}$ mass range of 15~GeV to $(m_H-m_Z)$, the upper bounds on $\delta^2\times \mathrm{BR}(Z_d \to 2\ell)$ are in the range $\sim(1.5$--$8.7)\times 10^{-5}$ as shown in Fig.~\ref{fig:delta}, assuming the same signal acceptances shown in Fig.~\ref{fig:effratios}. 

\section{$H \to Z_dZ_d \to 4\ell$}
\label{sec:ZdarkZdark}

\subsection{Search strategy}
\label{sec:stra}
\HZpZpllll{} candidate events are selected as discussed in Sec.~\ref{sec:evt}. The $Z$, $J/\psi$, $\Upsilon$ vetoes are applied as also discussed in Sec.~\ref{sec:evt}. Subsequently, the analysis exploits the small mass difference between the two
SFOS lepton pairs of the selected quadruplet to perform a counting experiment. After the small mass difference requirements between the SFOS lepton pairs, the 
estimated background contributions, coming from 
\HZZllll{} and \ZZllll{}, are small. These backgrounds are normalized with the theoretical calculations of their cross sections. The other backgrounds 
are found to be negligible. 
Since there is no significant excess, upper bounds on the signal strength, defined as the ratio of the \HZpZpllll{} rate normalized to the SM \HZZllll{} expectation are set as a function of the hypothesized $m_{Z_d}$.  
In a benchmark model where the SM is extended with a dark vector boson and a dark Higgs boson, the measured upper bounds on the signal strength are used
 to set limits on the branching ratio of $H\to Z_dZ_d$ and on the Higgs boson mixing parameter as a function of $m_{Z_d}$~\cite{Curtin:2014cca,Curtin:2013fra}.

\subsection{Event selection}
\label{sec:evt}

\begin{figure*}[!htbp]
\includegraphics[width=\textwidth]{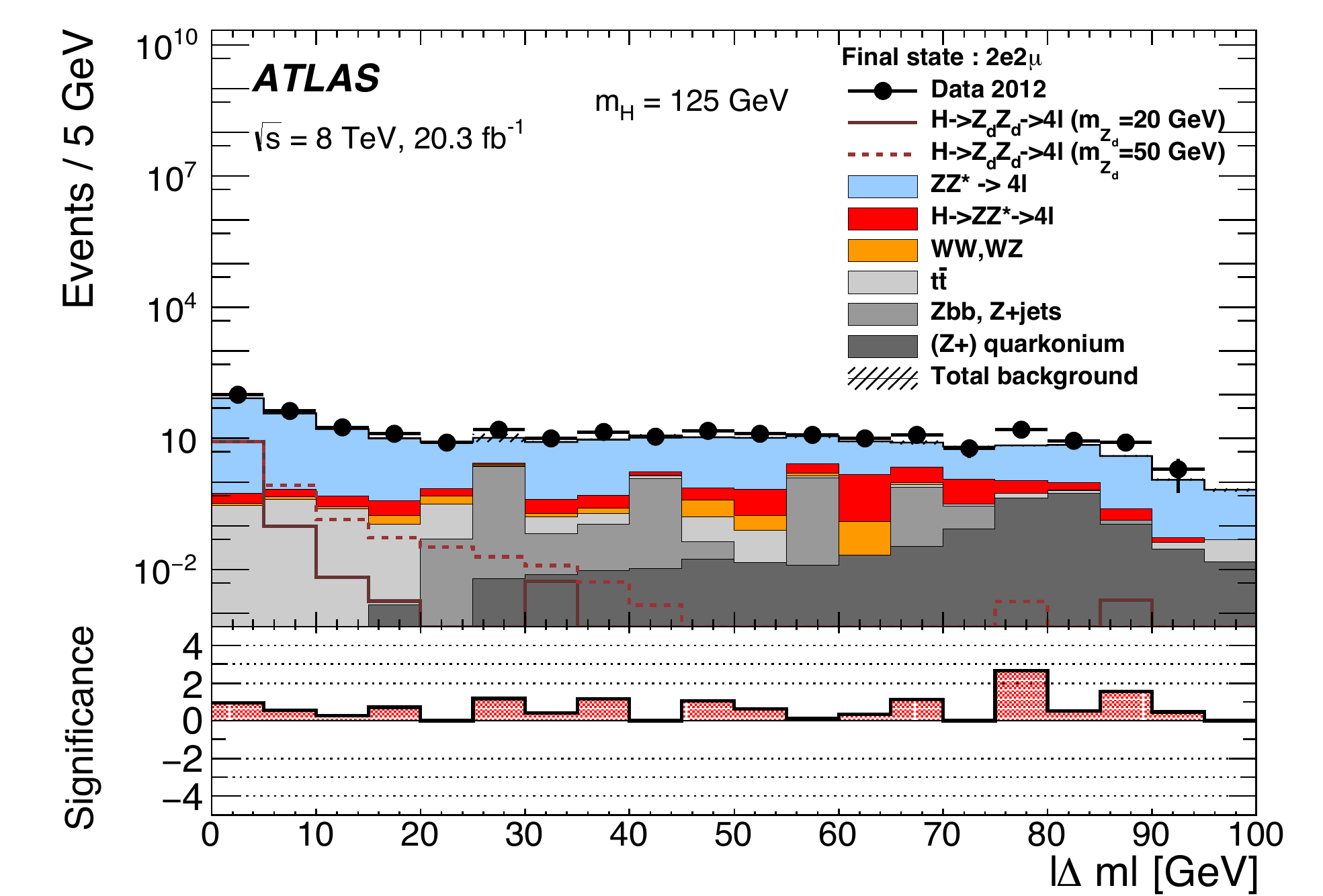}
\caption{Absolute mass difference between the two dilepton pairs, $\Delta m = |m_{12}-m_{34}|$ in the $2e2\mu$ channel, 
for $m_H=125$~GeV. %
The shaded area shows both the statistical and 
systematic uncertainties. The bottom plot shows the significance of the observed number of events in the data compared to the expected number of events 
from the 
backgrounds. These distributions are obtained after the impact parameter significance requirements.\label{ff:dmz_all_data}}
\end{figure*}

\begin{figure*}[htbp]
\includegraphics[width=\textwidth]{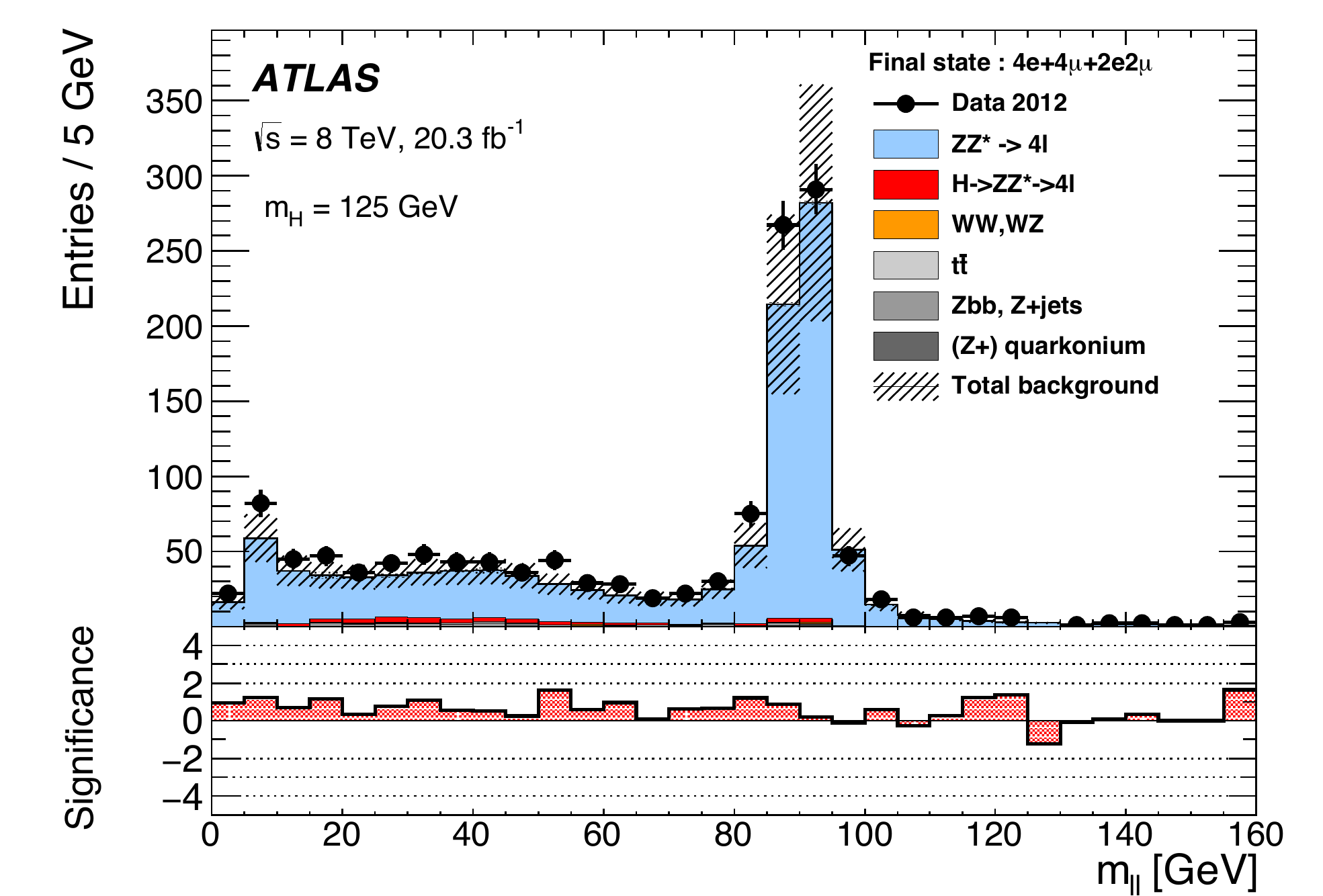}
\caption{Dilepton invariant mass, $m_{\ell\ell}\equiv m_{12}$ or $m_{34}$, in the combined $4e+2e2\mu+4\mu$ final state, for $m_H=125$~GeV. %
The shaded area shows both the statistical and systematic uncertainties. The bottom plots show the significance of the observed number of events in the data compared to the expected number of events from the backgrounds.  These distributions are obtained after the impact parameter significance requirement.\label{ff:m2l_all_data}}
\end{figure*}

\begin{figure*}[htbp]
\includegraphics[width=\textwidth]{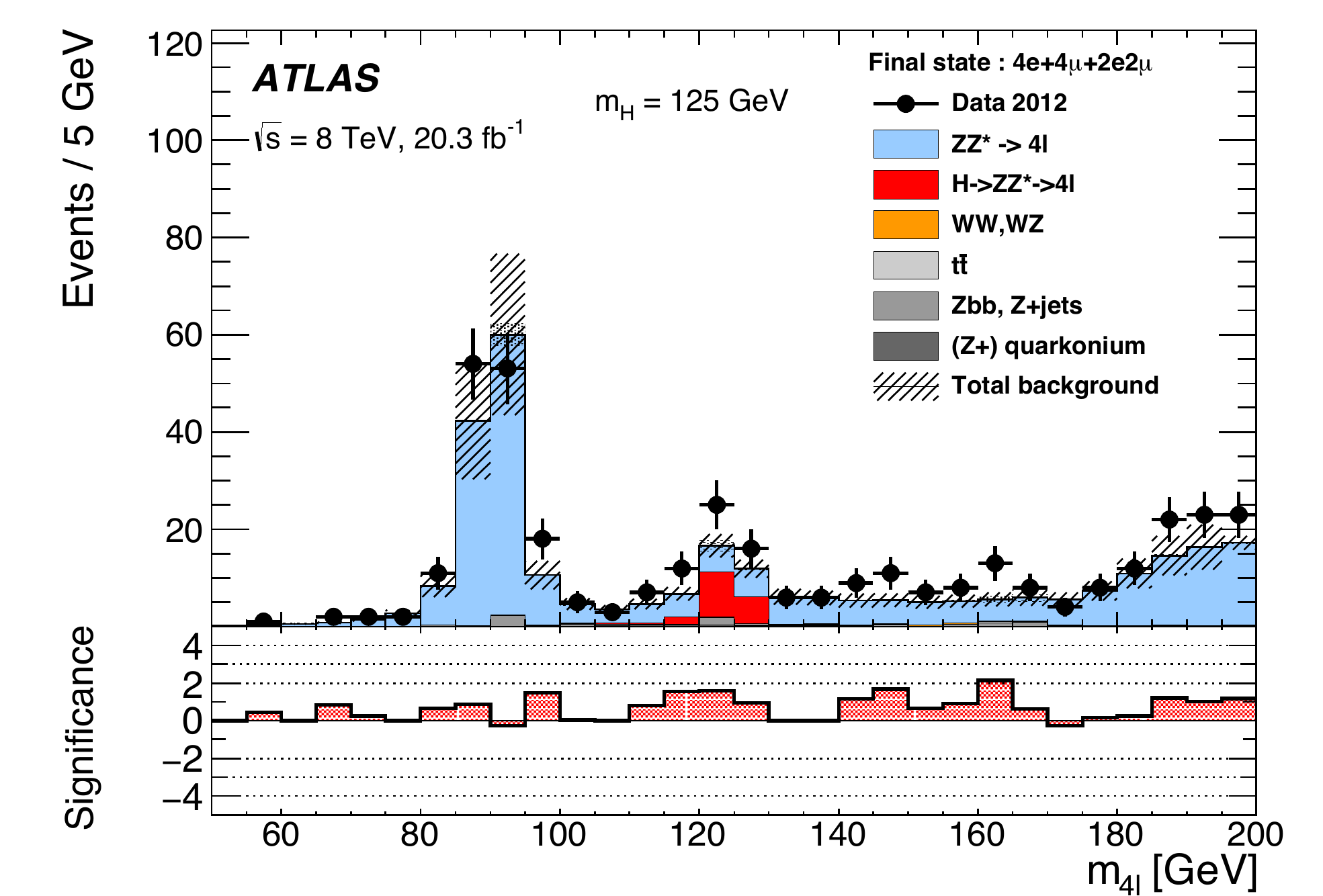}
\caption{Four-lepton invariant mass, in the combined $4e+2e2\mu+4\mu$ final state, for $m_H=125$~GeV. %
The shaded area shows both the statistical and systematic uncertainties. The bottom plots show the significance of the observed number of events in the data compared to the expected number of events from the backgrounds.  These distributions are obtained after the impact parameter significance requirement.\label{ff:m4l_all_data}}
\end{figure*}

For the \HZpZpllll{} search, unlike in the \HZZllll{} study~\cite{h4l2012}, there is no distinction between a 
primary pair (on-shell $Z$) and a secondary pair (off-shell $Z$), since both \Zp{} bosons are considered to be on-shell. 
Among all the different quadruplets, only one is selected by minimizing the mass difference $\Delta m = |m_{12} 
- m_{34}|$ where $m_{12}$ and $m_{34}$ are the invariant masses of the first and second pairs, respectively. The mass 
difference $\Delta m$ is expected to be minimal for the signal since the two dilepton systems should have invariant masses consistent with the same $m_{Z_d}$. No 
requirement is made on $\Delta m$; it is used only to select a unique quadruplet with the smallest $\Delta m$. Subsequently, isolation and impact parameter significance requirements are imposed on the leptons of the selected quadruplet as described in Ref.~\cite{moriond:2013t1}. Figure~\ref{ff:dmz_all_data} shows the minimal value of $\Delta m$  for the $2e2\mu$ final state after the impact parameter significance requirements. Similar distributions are found for the $4e$ and $4\mu$ final states. 
The dilepton and four-lepton invariant mass 
distributions are shown in Figs.~\ref{ff:m2l_all_data} and~\ref{ff:m4l_all_data} respectively for $m_{12}$ and $m_{34}$ combined.

For the \HZpZpllll{} search with hypothesized $m_{Z_d}$, after the impact parameter significance requirements on the selected leptons, four final requirements are applied:
\begin{enumerate}
\item[(1)] $115 < m_{4\ell} < 130~\GeV$ where $m_{4\ell}$ is the invariant mass of the four leptons in the quadruplet, consistent with the mass of the  discovered Higgs boson of about 125~GeV~\cite{Aad:2015zhl}.
\item[(2)] $Z$, $J/\psi$, and  $\Upsilon$ vetoes on all SFOS pairs in the selected quadruplet. The $Z$ veto discards the event
if either of the dilepton invariant masses is consistent with the $Z$-boson pole mass: $|m_{12}-m_Z|<10$~GeV or $|m_{34}-m_Z|<10$~GeV.  
For the $J/\psi$ and $\Upsilon$ veto, the dilepton invariant masses are required to be above $12$~GeV. This requirement suppresses
backgrounds with $Z$ bosons, $J/\psi$, and  $\Upsilon$. 
\item[(3)] the loose signal region requirement: $m_{12}<m_H/2$ and  $m_{34}< m_H/2$, where $m_H=125$~GeV. In the $H \to Z_d Z_d \to 4\ell$ search, the kinematic limit for on-shell $Z_d$ is $m_{Z_d} < m_H/2$. %
\item[(4)] the tight signal region requirement: $|m_{Z_d}-m_{12}|<\delta m$ and  $|m_{Z_d}-m_{34}|<\delta m$. The optimized values of the $\delta m$ requirements  are $5/3/4.5$~GeV for the $4e/4\mu/2e2\mu$ final states respectively (the $\delta m$ requirement varies with the hypothesized $m_{Z_d}$ but the impact of the variation is negligible). This requirement suppresses the backgrounds further by restricting the search region to within $\delta m$ of the hypothesized $m_{Z_d}$.  
\end{enumerate}
These requirements~(1)--(4) define the signal region (SR) of \HZpZpllll{} that is dependent on the hypothesized $m_{Z_d}$, and is essentially background-free, but contains small estimated background contributions 
from \HZZllll{} and \ZZllll{} processes as shown in Sec.~\ref{sec:results}.

\subsection{Background estimation}
\label{sec:backg}
\begin{figure*}[htbp]
\includegraphics[width=\textwidth]{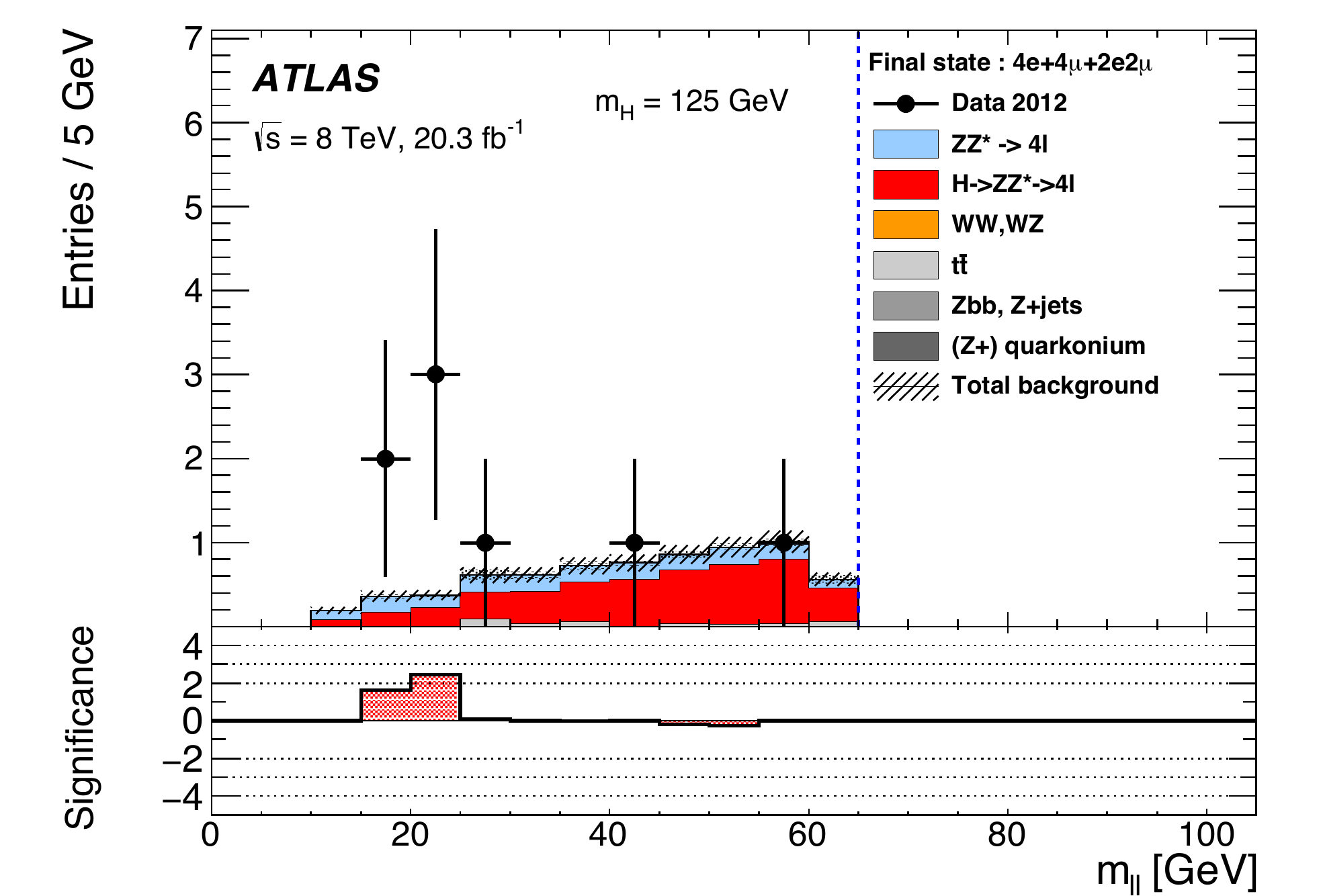}
\caption{Dilepton invariant mass, $m_{\ell\ell}\equiv m_{12}$ or $m_{34}$ after the loose signal region requirements described in Sec.~\ref{sec:evt} for the $4e$, $4\mu$ and, $2e2\mu$ final states combined,
for $m_H=125$~GeV. The data is represented by the black dots, and the backgrounds are represented by the filled histograms. The shaded area shows both the statistical and systematic uncertainties. The bottom plots show the significance of the observed data events compared to the expected number of events from the backgrounds.  The dashed vertical line is the kinematic limit ($m_{12}$ or $m_{34}<63$~GeV) of the loose signal region requirements as discussed in
Sec.~\ref{sec:evt}.}\label{ff:m2l_sr_data}
\end{figure*}

For the \HZpZpllll{} search, the main background contributions in the signal region come from the \HZZllll{} and \ZZllll{} 
processes.  These backgrounds are suppressed by the requirements of the tight signal region, as explained in Sec.~\ref{sec:evt}.
Other backgrounds with smaller contributions come from the $Z$+jets and $t\bar{t}$, $WW$ and $WZ$ processes as shown in Fig.~\ref{ff:m2l_sr_data}.
The \HZZllll{}, \ZZllll{}, $WW$ and $WZ$ backgrounds are estimated from simulation and normalized with theoretical
calculations of their cross sections. After applying the tight signal region requirements described in Sec.~\ref{sec:evt}, the $Z$+jets, $t\bar{t}$ 
and diboson backgrounds are negligible. In the case where the Monte Carlo calculation yields zero expected background events in the tight signal region,
an upper bound at 68\% CL on the expected events is estimated using 1.14 events~\cite{pdg}, scaled to the data luminosity and normalized to
the background cross section:

\begin{equation}
\label{eq:upperCount}
N_{\mathrm{background}} < L \times \sigma \times \left(\frac{1.14}{N_{\mathrm{tot}}}\right),
\end{equation} 
\noindent
where $L$ is the total integrated luminosity, $\sigma$ the cross section of the background process, and $N_{\mathrm{tot}}$ the total number of weighted events simulated for the background process.

\begin{figure*}[htbp]
\includegraphics[width=\textwidth]{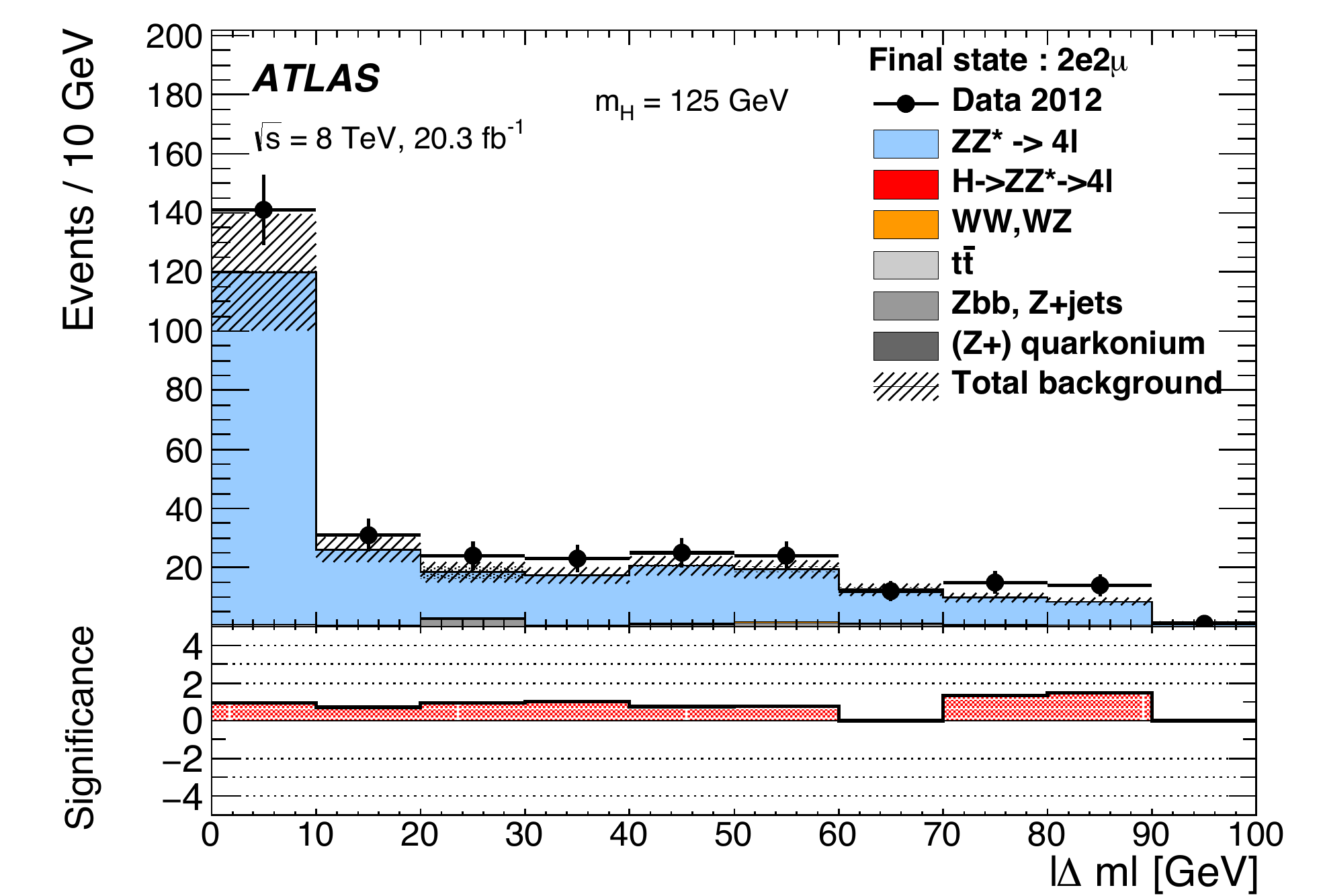}
\caption{The minimal absolute mass difference for the $2e2\mu$ final state.  Events are selected after the impact parameter significance requirement and
$m_{4\ell}\notin (115,130)$~$\GeV$. The shaded area shows both the statistical and systematic
uncertainties. The bottom plots show the significance of the measured number of events in the data compared to the estimated number of
events from the backgrounds.}
\label{CRa:dmz2e2mu}
\end{figure*}

To validate the background estimation, a signal depleted control region is defined by reversing the four-lepton invariant mass requirement with an $m_{4\ell}<115$~GeV or $m_{4\ell}>130$~GeV requirement. Good agreement between expectation and observation is found in this validation control region as shown in Fig.~\ref{CRa:dmz2e2mu}.

 \subsection{Systematic uncertainties}
 \label{sec:syst}

The systematic uncertainties on the theoretical calculations of the cross sections used in the event selection and identification efficiencies are taken into account. 
The effects of PDFs, $\alpha_{\mathrm{S}}$, and renormalization and factorization scale  uncertainties on the total inclusive cross sections for 
the Higgs production by ggF, VBF, $VH$ and $t\bar{t}H$ are obtained from
Refs.~\cite{LHCHiggsCrossSectionWorkingGroup:2011ti,LHCHiggsCrossSectionWorkingGroup:2012vm}. The renormalization, factorization scales and 
PDFs and $\alpha_{\mathrm{S}}$ uncertainties are applied to the $ZZ^*$ background estimates. 
The uncertainties due to the limited number of MC events in the $t\bar{t}$, $Z$+jets, $ZJ/\psi$, $Z\Upsilon$ and $WW/WZ$ background simulations are estimated as described in Sec.~\ref{sec:backg}.
The luminosity uncertainty~\cite{Aad:2013ucp} is applied to all signal yields, as well as to the background yields that are normalized with their theory cross 
sections. The detector systematic uncertainties due to uncertainties in the electron and muon identification efficiencies are estimated within the acceptance of the signal region requirements. There are 
several components to these uncertainties. For the muons, uncertainties in the  reconstruction and identification efficiency, and in the momentum resolution and scale, are included. For the 
electrons, uncertainties in the reconstruction and identification efficiency, the isolation and impact parameter significance requirements, the energy scale and energy resolution are considered.  The systematic uncertainties are summarized in Table~\ref{tab:ZdZd_syst}.
\begin{table}[!htbp]
  \begin{center}
    \begin{tabular}{lccc}
      \hline \hline
       \multicolumn{4}{c}{Systematic Uncertainties (\%)} \\
      \hline \hline
      Source &  4$\mu$ & 4e & 2e2$\mu$ \\
      \hline
       Electron Identification &  -- & 6.7 & 3.2 \\
       Electron Energy Scale   &  -- & 0.8 & 0.3 \\
       Muon Identification     & 2.6 & -- & 1.3 \\
       Muon Momentum Scale     & 0.1 & -- & 0.1 \\
       Luminosity              & 2.8 & 2.8 & 2.8 \\
       ggF QCD                 & 7.8 & 7.8 & 7.8 \\
       ggF PDFs and $\alpha_{\mathrm{S}}$      & 7.5 & 7.5 & 7.5 \\
       $ZZ^*$ Normalization    & 5.0 & 5.0 & 5.0 \\
      \hline \hline
    \end{tabular}
  \caption{The relative systematic uncertainties on the event yields in the $H\to Z_dZ_d \to 4\ell$  search.}
  \label{tab:ZdZd_syst}
  \end{center}
\end{table}

\subsection{Results and interpretation}
\label{sec:results}

\begin{figure*}[htbp]
\includegraphics[width=\textwidth]{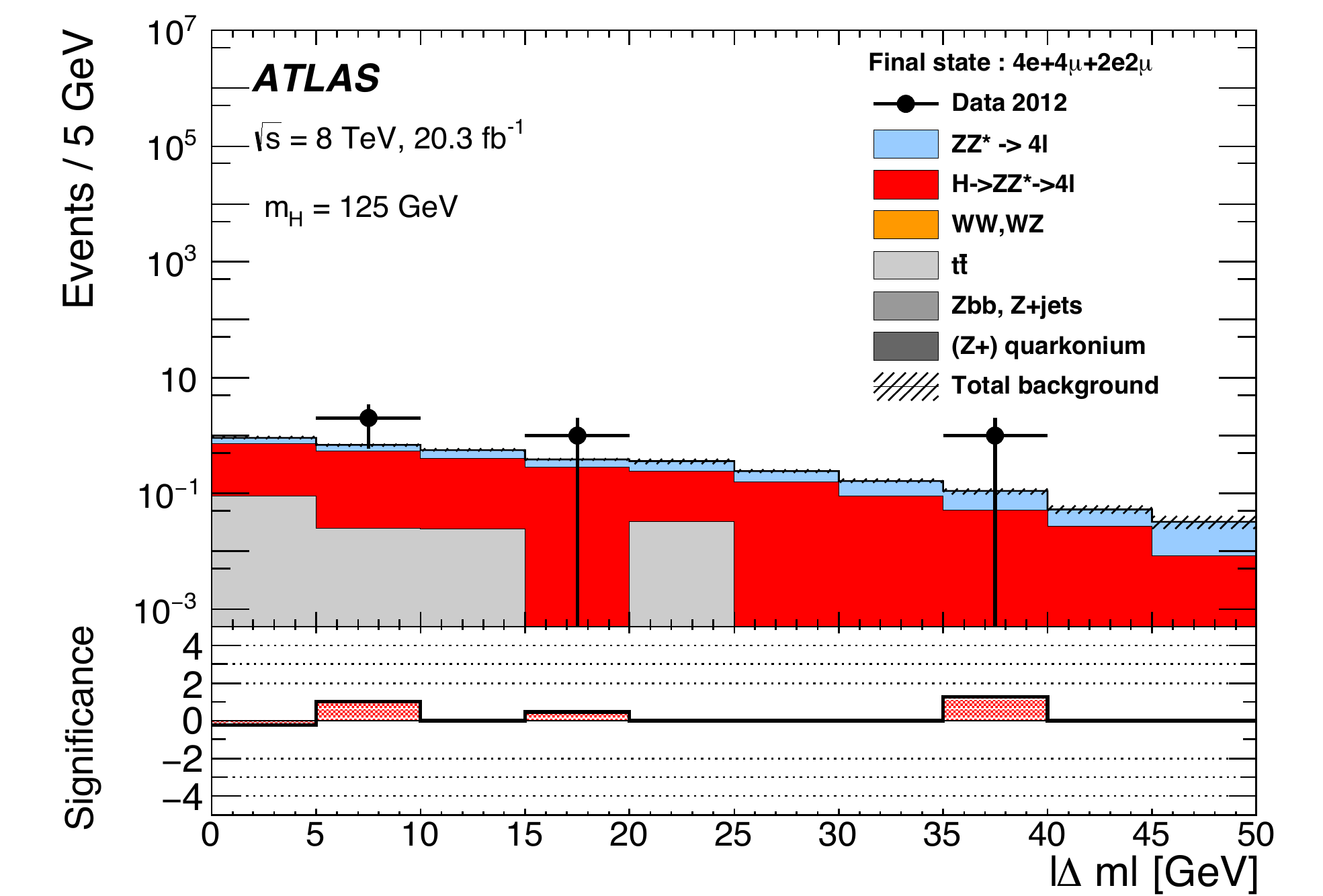}
\caption{Absolute mass difference, $\Delta m=|m_{12}-m_{34}|$ after the loose signal region requirements described in Sec.~\ref{sec:evt} for the $4e$, $4\mu$ and, $2e2\mu$ final states combined,
for $m_H=125$~GeV. The data is represented by the black dots, and the backgrounds are represented by the filled histograms. The shaded area shows both the statistical and systematic uncertainties. The bottom plots show the significance of the observed data events compared to the expected number of events from the backgrounds.}\label{ff:dm_sr_data}
\end{figure*}

Figures~\ref{ff:m2l_sr_data} and~\ref{ff:dm_sr_data} show the distributions of the dilepton invariant mass (for $m_{12}$ and $m_{34}$ combined) and the absolute mass difference $\Delta m=|m_{12}-m_{34}|$ after the loose signal region
requirements. 
Four data events pass the loose signal region requirements, one in the $4e$ channel, two in the $4\mu$ channel and one in the $2e2\mu$ channel. Two of these four events pass the tight signal region requirements: the event in the $4e$ channel and one of the events in the $4\mu$ channel.  The event in the $4e$ channel has dilepton masses of 21.8~GeV and 28.1~GeV as shown in Fig.~\ref{ff:m2l_sr_data}, and is consistent with a $Z_d$ mass in the range $23.5 \leq m_{Z_d} \leq 26.5$~GeV. %
For the event in the $4\mu$ channel that passes the tight signal region requirements, the dilepton invariant masses are 23.2~GeV and 18.0~GeV as shown in Fig.~\ref{ff:m2l_sr_data}, and they are consistent with a $Z_d$ mass in the range $20.5 \leq m_{Z_d} \leq 21.0$~GeV. %
In the $m_{Z_d}$ range of 15 to 30~GeV where four data events pass the loose signal region requirements, histogram interpolation~\cite{atlasTool:1999} is used in steps of
0.5~GeV to obtain the signal acceptances and efficiencies at the hypothesized $m_{Z_d}$.
 The expected numbers of signal, background and data events, after applying the tight signal region requirements, are shown in Table~\ref{tab:final_events}.
\begin{table*}[tb]
\begin{center}
\begin{tabular}{cccc}
\hline
\hline
                                       Process & $                4e$ & $              4\mu$ & $            2e2\mu$             \\ \hline\hline
                                      \HZZllll & $ (1.5 \pm 0.3 \pm 0.2)\times 10^{-2}$ & $ (1.0 \pm 0.3 \pm 0.3)\times 10^{-2}$ & $(2.9 \pm 1.0  \pm 2.0)\times 10^{-3} $ \\
                                       \ZZllll & $ (7.1 \pm 3.6 \pm 0.5)\times 10^{-4}$  & $(8.4 \pm 3.8 \pm  0.5)\times 10^{-3}$  & $(9.1 \pm 3.6 \pm 0.6)\times 10^{-3}$   \\     
                                       $WW,WZ$ & $   <0.7\times 10^{-2}$   & $  <0.7\times 10^{-2}$   & $<0.7\times 10^{-2}$ 		\\
                                    $t\bar{t}$ & $   <3.0\times 10^{-2}$   & $  <3.0\times 10^{-2}$   & $ <3.0\times 10^{-2}$	\\
                                    $Zbb, Z$+jets & $ <0.2\times 10^{-2}$  & $              <0.2\times 10^{-2}$  & $   <0.2\times 10^{-2}$    \\
                                    $ZJ/\psi$ and $Z\Upsilon$ & $ <2.3\times 10^{-3}$   & $        <2.3\times 10^{-3}$ & $    <2.3\times 10^{-3}$  \\
\hline
                             Total background & $ <5.6\times 10^{-2}$  & $ < 5.9\times 10^{-2}$  & $ <5.3\times 10^{-2}$        \\
                                          Data & $   1$  & $   0$  & $   0$               \\
\hline\hline
                                 \HZZllll & $ (1.2 \pm 0.3 \pm 0.2)\times 10^{-2}$  & $ (5.8 \pm 2.0 \pm 2.0)\times 10^{-3}$  & $(2.6 \pm 1.0  \pm 0.2)\times 10^{-3} $ \\
                                 \ZZllll & $ (3.5 \pm 2.0 \pm 0.2)\times 10^{-3}$  & $(4.1 \pm 2.7 \pm  0.2)\times 10^{-3}$  & $(2.0 \pm 0.6 \pm 0.1)\times 10^{-2}$   \\
                                       $WW,WZ$ & $   <0.7\times 10^{-2}$   & $  <0.7\times 10^{-2}$   & $<0.7\times 10^{-2}$               \\
                                    $t\bar{t}$ & $   <3.0\times 10^{-2}$   & $  <3.0\times 10^{-2}$   & $ <3.0\times 10^{-2}$      \\
                                    $Zbb, Z$+jets & $ <0.2\times 10^{-2}$  & $              <0.2\times 10^{-2}$  & $   <0.2\times 10^{-2}$    \\
                                    $ZJ/\psi$ and $Z\Upsilon$ & $ <2.3\times 10^{-3}$   & $        <2.3\times 10^{-3}$ & $    <2.3\times 10^{-3}$  \\
\hline
                             Total background & $ <5.3\times 10^{-2}$  & $ <5.1\times 10^{-2}$  & $ <6.4\times 10^{-2}$        \\
                                          Data & $   0$  & $   1$  & $   0$               \\

\hline\hline

\end{tabular}
\caption{The expected and observed numbers of events in the tight signal region of the $H\to Z_dZ_d\to 4\ell$ search for each of the three final states, for 
the hypothesized  mass $m_{Z_d}=25$~GeV and 20.5~GeV. Statistical and systematic uncertainties are given respectively for the signal and the background expectations. One 
event in data passes all the selections in the $4e$ channel and is consistent with $23.5 \leq m_{Z_d} \leq 26.5$~GeV. One other data event passes all the selections in the $4\mu$ 
channel and is consistent with $20.5 \leq m_{Z_d} \leq 21.0$~GeV. The \HZZllll{} numbers are summed over the ggF, VBF, $ZH$, $WH$ and $t\bar{t}H$ processes. } 
\label{tab:final_events}
\end{center}
\end{table*}

\begin{figure*}[!htbp]
\includegraphics[width=\textwidth]{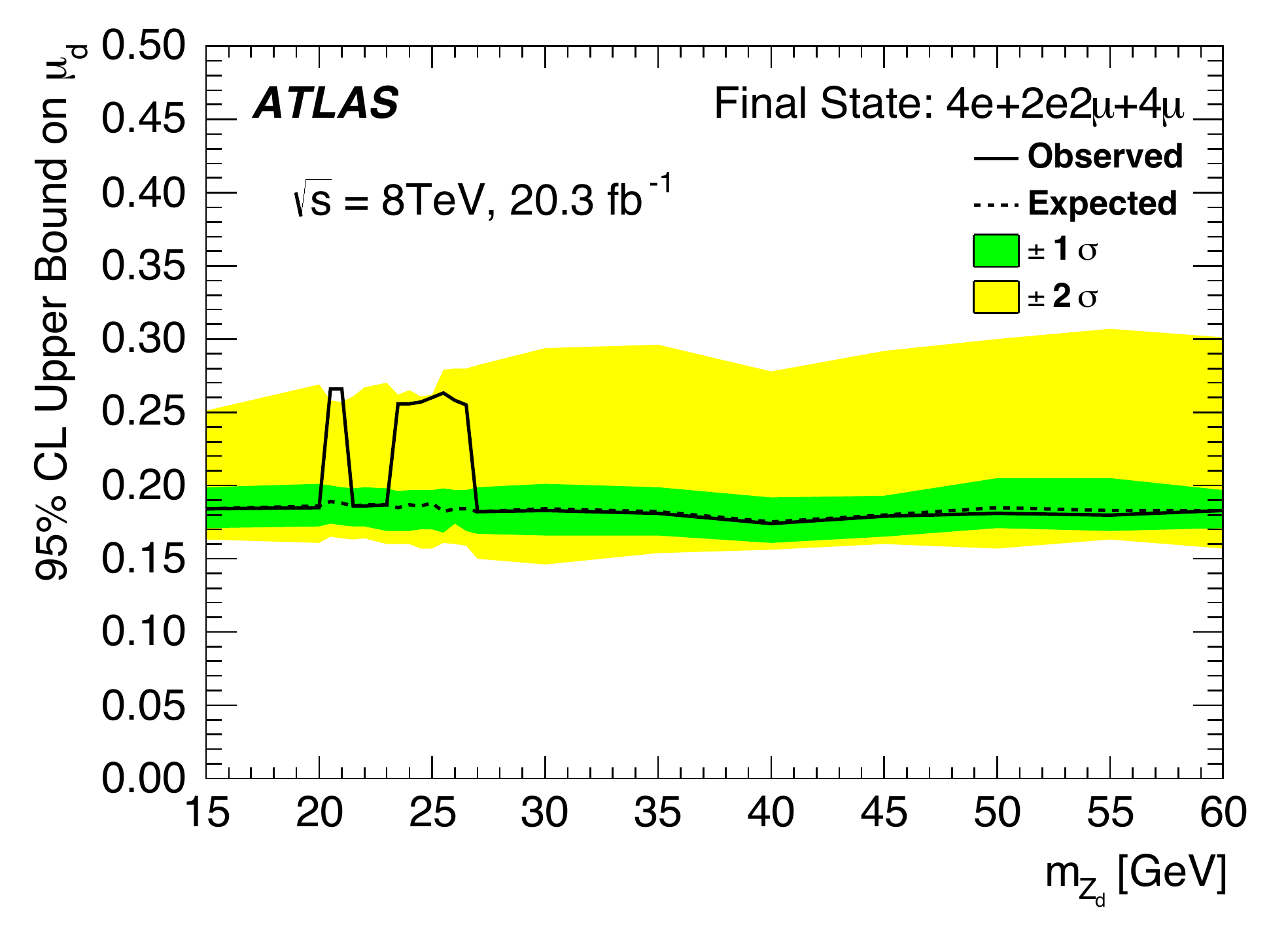}
\caption{The 95\% confidence level upper bound on the signal strength $\mu_d = \frac{\sigma \times \mathrm{BR}(\HZpZpllll) }{[\sigma \times \mathrm{BR} (\HZZllll)]_{\mathrm{SM}}}$ of \HZpZpllll{} in the combined $4e+2e2\mu+4\mu$ final state, for $m_H=125$~GeV.
The $\pm 1\sigma$ and $\pm 2\sigma$ expected exclusion regions are indicated in green and yellow, respectively.\label{fig:HZpZp}}
\end{figure*}

\begin{figure*}[!htbp]
\includegraphics[width=\textwidth]{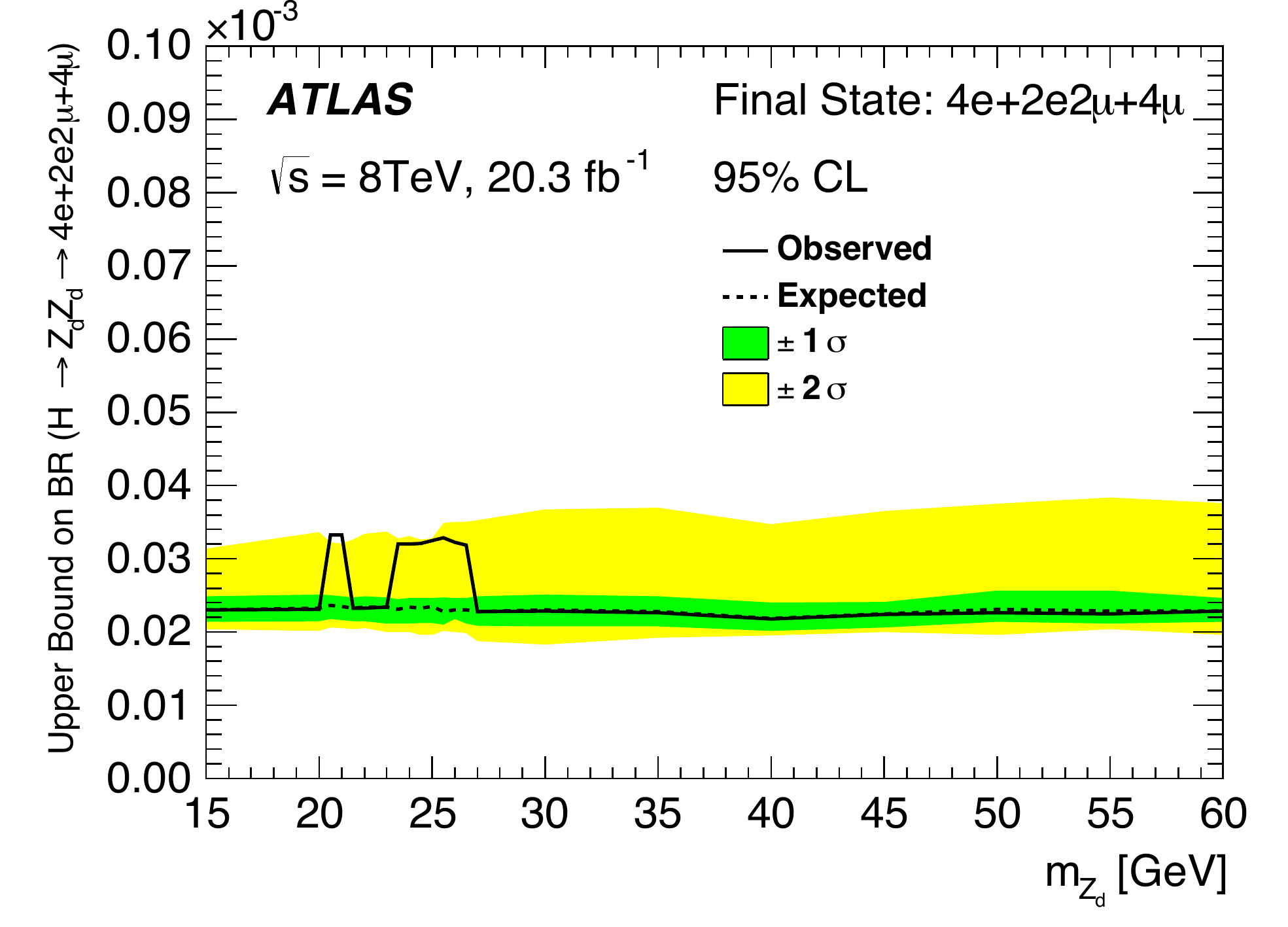}
\caption{The 95\% confidence level upper bound on the branching ratio of \HZpZpllll{} as a function of $m_{Z_d}$, in the combined $4e+2e2\mu+4\mu$ final state, for $m_H=125$~GeV.
The $\pm 1\sigma$ and $\pm 2\sigma$ expected exclusion regions are indicated in green and yellow, respectively.\label{fig:HZpZp4l}}
\end{figure*}

For each $m_{Z_d}$, in the absence of any significant excess of events consistent with the signal hypothesis, the upper limits are computed from a maximum-likelihood fit to the numbers of data and expected signal and background events in
the tight signal regions, following the $CL_s$ modified frequentist formalism~\cite{CLs,Cousins:1991} with the profile-likelihood test statistic~\cite{2011EPJC...71.1554C,eCowan}.
The nuisance parameters associated to the systematic uncertainties described in Sec.~\ref{sec:syst} are profiled. The parameter of interest in the fit is the signal
strength $\mu_{d}$ defined as the ratio of the \HZpZpllll{} rate relative to the SM \HZZllll{} rate:

\begin{equation}
\label{eq:mudark}
\mu_{d} = \frac{\sigma \times \mathrm{BR}(\HZpZpllll) }{[\sigma \times \mathrm{BR} (\HZZllll)]_{\mathrm{SM}}}.
\end{equation}
\noindent
The systematic uncertainties in the electron and muon identification efficiencies, renormalization and factorization scales and PDF are 100\%
correlated between the signal and backgrounds. Pseudoexperiments are used to compute the 95\% CL upper bound $\mu_d$ 
in each of the final states and their combination, and for each of the hypothesized $m_{Z_d}$.
The 95\% confidence-level upper bounds on the \HZpZpllll{} rates are shown in Fig.~\ref{fig:HZpZp} relative to the SM Higgs boson process  \HZZllll{} as a function of
the hypothesized $m_{Z_d}$ for the combination of the three final states $4e$, $2e2\mu$ and $4\mu$.
Assuming the SM Higgs boson production cross section and using $\mathrm{BR} (\HZZllll)_{\mathrm{SM}} = 1.25\times 10^{-4}$~\cite{LHCHiggsCrossSectionWorkingGroup:2011ti,LHCHiggsCrossSectionWorkingGroup:2012vm}, upper bounds on the branching ratio of \HZpZpllll{} can be obtained from Eq.~(\ref{eq:mudark}), as shown in Fig.~\ref{fig:HZpZp4l}.

\begin{figure*}[!htbp]
\includegraphics[width=\textwidth]{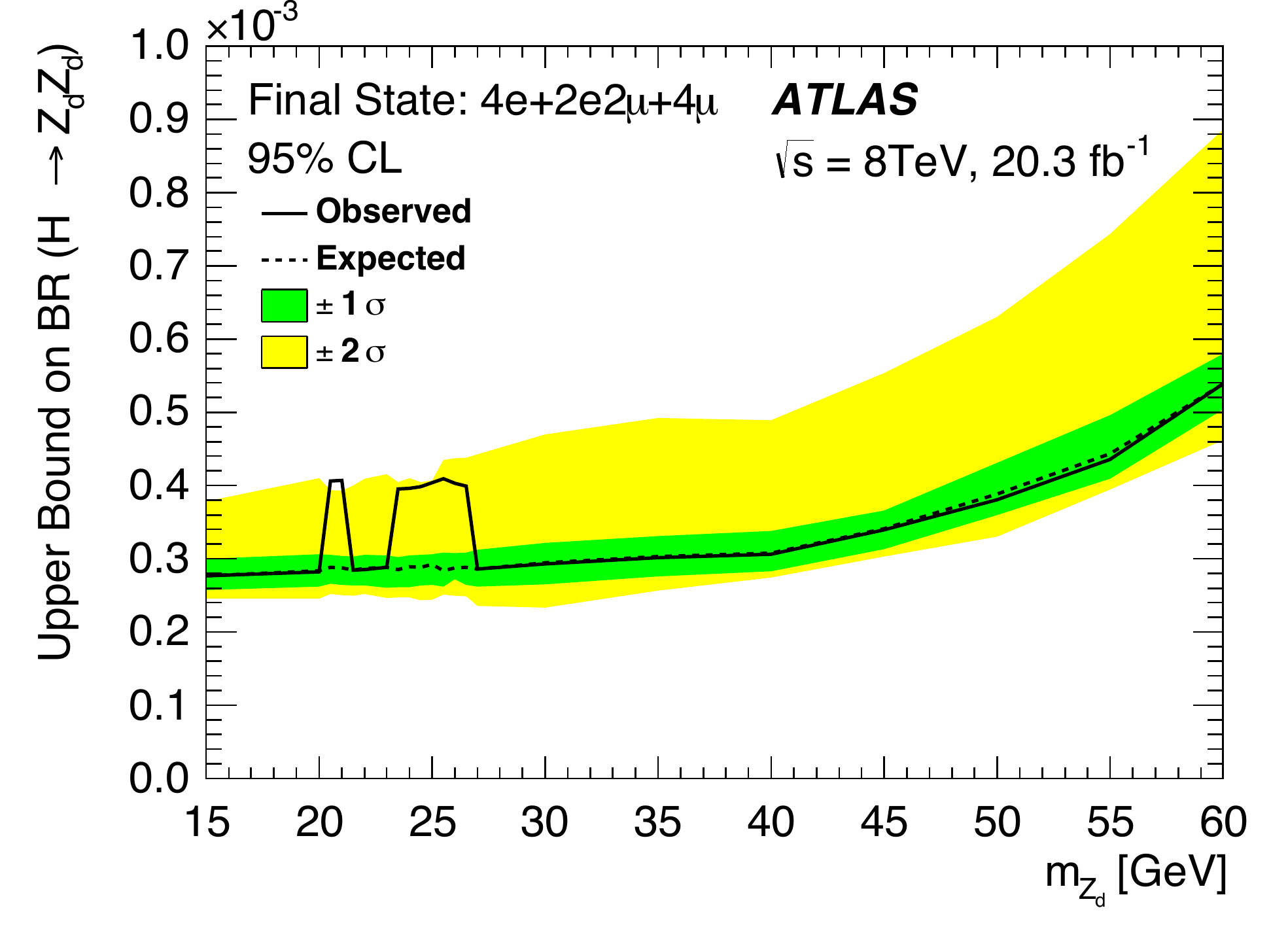}
\caption{The 95\% confidence level upper bound on the branching ratio of $H\rightarrow Z_dZ_d$ in the combined $4e+2e2\mu+4\mu$ final state, for $m_H=125$~GeV.
The $\pm 1\sigma$ and $\pm 2\sigma$ expected exclusion regions are indicated in green and yellow, respectively.\label{fig:BrZdZd}}
\end{figure*}

The simplest benchmark model is the SM plus a dark vector boson and a dark Higgs boson as discussed in  Refs.~\cite{Curtin:2013fra,gopalakrishna2008higgs}, where the branching ratio of 
$Z_d\rightarrow\ell\ell$ is given as a function of $m_{Z_d}$. This can be used to convert the measurement of the upper bound on
the signal strength $\mu_{d}$ into an upper bound on the branching ratio $\mathrm{BR}(H\rightarrow Z_dZ_d)$ assuming the SM Higgs boson production
cross section. Figure~\ref{fig:BrZdZd} shows the 95\% CL upper limit on the branching ratio of $H\rightarrow Z_dZ_d$ as a function of $m_{Z_d}$ using the combined $\mu_{d}$ of the three final states. The weaker bound at higher $m_{Z_d}$ is due to the fact that the branching ratio $Z_d 
\rightarrow\ell\ell$ drops slightly at higher $m_{Z_d}$~\cite{Curtin:2013fra} as other decay channels become accessible.
\begin{figure*}[!htbp]
\includegraphics[width=\textwidth]{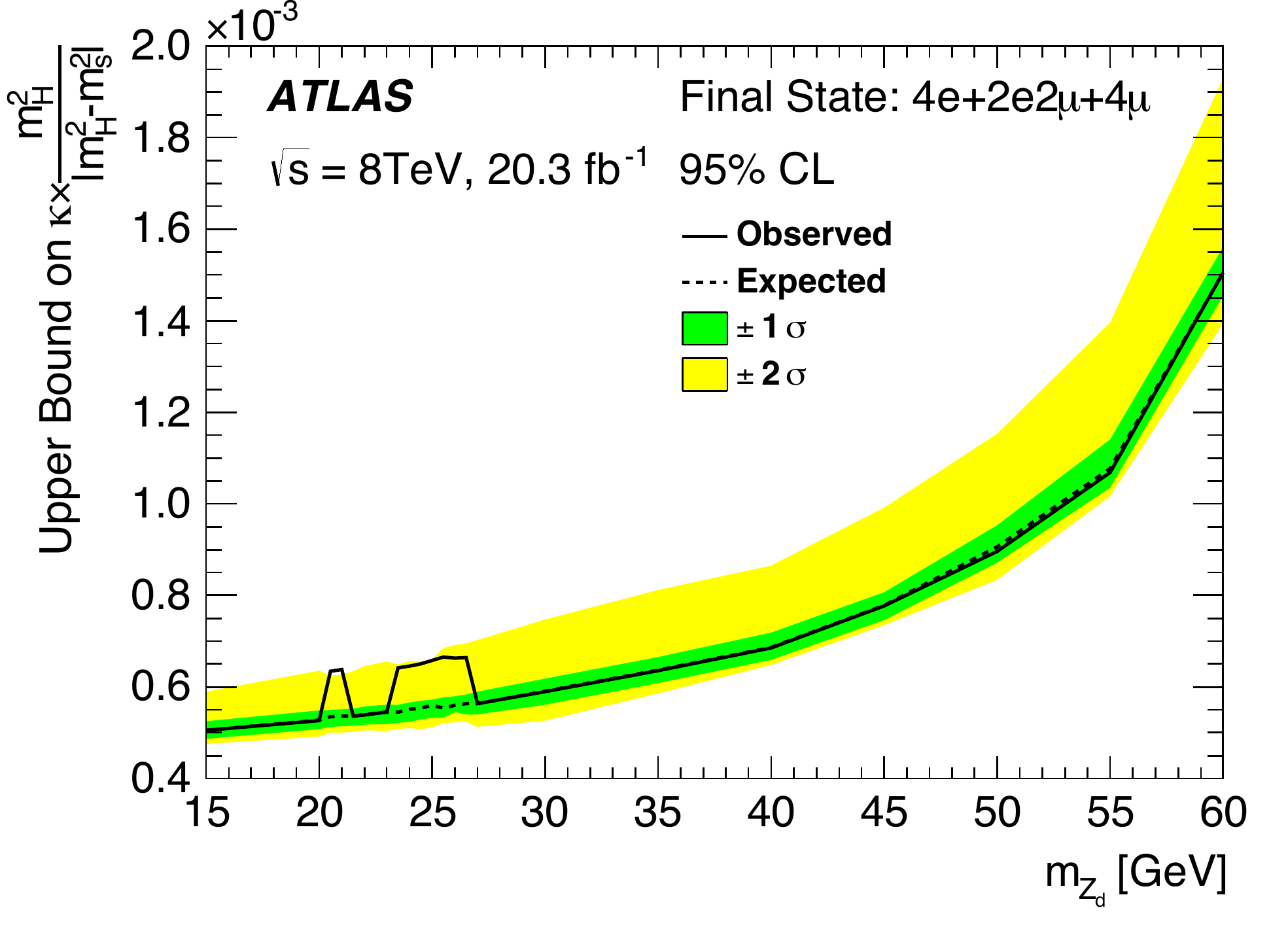}
\caption{The 95\% confidence level upper bound on the Higgs mixing parameter $\kappa \times m_H^2/|m_H^2-m_S^2|$  as a function of $m_{Z_d}$, in the combined $4e+2e2\mu+4\mu$ final state, for $m_H=125$~GeV.
The $\pm 1\sigma$ and $\pm 2\sigma$ expected exclusion regions are indicated in green and yellow, respectively.\label{fig:kappa}}
\end{figure*}
The $H\rightarrow Z_dZ_d$ decay can be used to obtain a $m_{Z_d}$-dependent limit on an Higgs mixing parameter $\kappa'$~\cite{Curtin:2013fra}:

\begin{equation}
\label{eq:kappa}
\kappa' = \kappa \times \frac{m_H^2}{|m_H^2-m_S^2|},
\end{equation}
\noindent
where $\kappa$ is the size of the Higgs portal coupling and $m_S$ is the mass of the dark Higgs boson. The partial width of $H\rightarrow Z_dZ_d$ is given
in terms of $\kappa$~\cite{Curtin:2014cca}. In the regime where the Higgs mixing parameter dominates ($\kappa \gg \epsilon$), $m_S > m_H/2$, $m_{Z_d} < m_H/2$ and $H \to Z_dZ^{*}\to 4\ell$
is negligible, the only relevant decay is
$H\rightarrow Z_dZ_d$. Therefore the partial width $\Gamma (H\rightarrow Z_dZ_d)$ can be written as:

\begin{equation}
\label{eq:gamma}
\Gamma (H\rightarrow Z_dZ_d) = \Gamma_{\mathrm{SM}}\frac{\mathrm{BR}(H\rightarrow Z_dZ_d)}{1-\mathrm{BR}(H\rightarrow Z_dZ_d)}.
\end{equation}
\noindent
The Higgs portal coupling parameter $\kappa$ is obtained using Eq.~(53) of Ref.~\cite{Curtin:2013fra} or Table~2 of Ref.~\cite{Curtin:2014cca}:

\begin{equation}
\label{eq:kappa2}
\kappa^2 = \frac{\Gamma_{\mathrm{SM}}}{f(m_{Z_d})}\frac{\mathrm{BR}(H\rightarrow Z_dZ_d)}{1-\mathrm{BR}(H\rightarrow Z_dZ_d)},
\end{equation}
\noindent
where
\begin{align}
\label{eq:fZdZd}
f(m_{Z_d}) = & \frac{v^2}{32\pi m_H} \times \sqrt{1-\frac{4m^2_{Z_d}}{m^2_H}} \times \nonumber \\
& \frac{\left(m^2_H+2m^2_{Z_d}\right)^2-8\left(m^2_H-m^2_{Z_d}\right)m^2_{Z_d}}{\left(m^2_H-m^2_S\right)^2}.
\end{align}
\noindent

Figure~\ref{fig:kappa} shows the upper bound on the effective Higgs mixing parameter as a 
function of $m_{Z_d}$: for $m_H/2 < m_S < 2m_H$, this would correspond to an upper bound on the Higgs portal coupling in the range $\kappa \sim (1$--$10)\times10^{-4}$.

An interpretation for $H \to Z_dZ_d$ is not done in the $Z$-$Z_d$ mass mixing scenario described in Refs.~\cite{Davoudiasl:2013aya,BNL2012dark} since in this
model the rate of $H \to Z_dZ_d$ is highly suppressed relative to that of $H \to ZZ_d$.

\section{Conclusions}
\label{sec:conc}
Two searches for an exotic gauge boson $\Zdark$ that
couples to the discovered SM Higgs boson at a mass around 125~GeV in four-lepton events are presented, using the ATLAS detector at the LHC. 

The $\HZdfl$ analysis uses the
events resulting from Higgs boson decays to four leptons to search for
an exotic gauge boson $\Zdark$, by examining the $\mthreefour$ mass distribution.  
The results obtained in this search cover the exotic gauge boson mass range of $15 < m_{Z_d} < 55$~GeV, 
 and are based on proton-proton collisions data at $\sqrt{s}$ = 8~TeV with an integrated 
luminosity of 20.7~$\ifb$. Observed and expected exclusion limits on 
the branching ratio of $\HZdfl$ relative to $H \to 4\ell$ are estimated for the combination of all the final states.  
For relative branching ratios above 0.4 (observed) and 0.2 (expected), the entire mass range of $15 < m_{Z_d} < 55$~GeV is excluded 
at 95\% CL.  %
Upper bounds at 95\% CL on the branching ratio of $\HZdfl$ are set in the range $(1$--$9)\times10^{-5}$ 
for $15 < m_{Z_d} < 55$~GeV, assuming the SM branching ratio of $\HZsfl$. 

The \HZpZpllll{} search covers the exotic gauge boson mass
range from 15~GeV up to the kinematic limit of $m_H/2$. An integrated luminosity of 20.3~\ifb\ at 8~TeV is 
used in this search. One data event is observed to  pass all the signal region selections in the $4e$ channel, and has 
dilepton invariant masses of 21.8~GeV and 28.1~GeV. %
This $4e$ event is consistent with a $Z_d$ mass in the range $23.5 < m_{Z_d} < 26.5$~GeV. Another data event is observed to pass all the signal region selections in the $4\mu$ channel, and has dilepton invariant masses of 23.2~GeV and 18.0~GeV.%
This $4\mu$ event is consistent with a $Z_d$ mass in the range $20.5 < m_{Z_d} < 21.0$~GeV. In the absence of a significant excess, upper bounds on the signal strength (and thus on the cross section times branching
ratio) are set for the mass range of $15 < m_{Z_d} < 60$~GeV using the combined $4e$, $2e2\mu$, $4\mu$ final states. 

Using a simplified model where the SM is extended with the addition of an exotic gauge boson and a dark Higgs boson,
upper bounds on the gauge kinetic mixing parameter $\epsilon$ (when $\epsilon$~$\gg$~$\kappa$), 
are set in the range $(4$--$17)\times10^{-2}$ at 95\% CL, assuming the SM branching ratio of $\HZsfl$, for $15 < m_{Z_d} < 55$~GeV. 
Assuming the SM Higgs production cross section, upper bounds on the branching ratio of $H\rightarrow Z_dZ_d$, as well as on the Higgs portal coupling parameter $\kappa$ are set in the range  $(2$--$3)\times10^{-5}$ and $(1$--$10)\times 10^{-4}$ respectively at 95\% CL, for $15 < m_{Z_d} < 60$~GeV.

Upper bounds on the effective mass-mixing parameter $\delta^2\times \mathrm{BR}(Z_d \to \ell\ell)$, resulting from the $U(1)_d$ gauge symmetry, are also set using the branching ratio measurements in the $H \to ZZ_d \to 4\ell$ search, and are in the range $(1.5-8.7)\times 10^{-5}$ for $15<m_{Z_d} < 35$~GeV.

\section*{Acknowledgments}

We thank CERN for the very successful operation of the LHC, as well as the
support staff from our institutions without whom ATLAS could not be
operated efficiently.

We acknowledge the support of ANPCyT, Argentina; YerPhI, Armenia; ARC,
Australia; BMWFW and FWF, Austria; ANAS, Azerbaijan; SSTC, Belarus; CNPq and FAPESP,
Brazil; NSERC, NRC and CFI, Canada; CERN; CONICYT, Chile; CAS, MOST and NSFC,
China; COLCIENCIAS, Colombia; MSMT CR, MPO CR and VSC CR, Czech Republic;
DNRF, DNSRC and Lundbeck Foundation, Denmark; EPLANET, ERC and NSRF, European Union;
IN2P3-CNRS, CEA-DSM/IRFU, France; GNSF, Georgia; BMBF, DFG, HGF, MPG and AvH
Foundation, Germany; GSRT and NSRF, Greece; ISF, MINERVA, GIF, I-CORE and Benoziyo Center,
Israel; INFN, Italy; MEXT and JSPS, Japan; CNRST, Morocco; FOM and NWO,
Netherlands; BRF and RCN, Norway; MNiSW and NCN, Poland; GRICES and FCT, Portugal; MNE/IFA, Romania; MES of Russia and ROSATOM, Russian Federation; JINR; MSTD,
Serbia; MSSR, Slovakia; ARRS and MIZ\v{S}, Slovenia; DST/NRF, South Africa;
MINECO, Spain; SRC and Wallenberg Foundation, Sweden; SER, SNSF and Cantons of
Bern and Geneva, Switzerland; NSC, Taiwan; TAEK, Turkey; STFC, the Royal
Society and Leverhulme Trust, United Kingdom; DOE and NSF, United States of
America.

The crucial computing support from all WLCG partners is acknowledged
gratefully, in particular from CERN and the ATLAS Tier-1 facilities at
TRIUMF (Canada), NDGF (Denmark, Norway, Sweden), CC-IN2P3 (France),
KIT/GridKA (Germany), INFN-CNAF (Italy), NL-T1 (Netherlands), PIC (Spain),
ASGC (Taiwan), RAL (UK) and BNL (USA) and in the Tier-2 facilities
worldwide.

\clearpage
\printbibliography

\clearpage
\begin{flushleft}
{\Large The ATLAS Collaboration}

\bigskip

G.~Aad$^{\rm 85}$,
B.~Abbott$^{\rm 113}$,
J.~Abdallah$^{\rm 151}$,
O.~Abdinov$^{\rm 11}$,
R.~Aben$^{\rm 107}$,
M.~Abolins$^{\rm 90}$,
O.S.~AbouZeid$^{\rm 158}$,
H.~Abramowicz$^{\rm 153}$,
H.~Abreu$^{\rm 152}$,
R.~Abreu$^{\rm 30}$
% ...
\bigskip
\\
% ...
$^{11}$ Institute of Physics, Azerbaijan Academy of Sciences, Baku, Azerbaijan\\
% ...
$^{30}$ CERN, Geneva, Switzerland\\
% ...
$^{85}$ CPPM, Aix-Marseille Universit{\'e} and CNRS/IN2P3, Marseille, France\\
% ...
$^{90}$ Department of Physics and Astronomy, Michigan State University, East Lansing MI, United States of America\\
% ...
$^{107}$ Nikhef National Institute for Subatomic Physics and University of Amsterdam, Amsterdam, Netherlands\\
% ...
$^{113}$ Homer L. Dodge Department of Physics and Astronomy, University of Oklahoma, Norman OK, United States of America\\
% ...
$^{151}$ Institute of Physics, Academia Sinica, Taipei, Taiwan\\
$^{152}$ Department of Physics, Technion: Israel Institute of Technology, Haifa, Israel\\
$^{153}$ Raymond and Beverly Sackler School of Physics and Astronomy, Tel Aviv University, Tel Aviv, Israel\\
% ...
$^{158}$ Department of Physics, University of Toronto, Toronto ON, Canada\\
% ...
\end{flushleft}

\end{document}